\documentclass[aps,prd,amsmath,longbibliography]{revtex4-1}

\usepackage[dvips]{graphicx}
\usepackage{latexsym,epsfig,bm,psfrag,subfigure}
\usepackage{color}         
\usepackage[bookmarksnumbered,bookmarksopen,colorlinks,citecolor=blue,linkcolor=blue]{hyperref} %
\def\xX{x_X}

\newcommand{\eq}[1]{Eq.~(\ref{#1})}

\newcommand{\bfzeta}{\mbox{\boldmath$\zeta$}}
\def\bea{\begin{eqnarray}}
\def\eea{\end{eqnarray}}\newcommand{\bfk}{{\bf k}}\newcommand{\bfq}{{\bf q}}
\def\g{\gamma}\def\m{\mu}\def\ve{\varepsilon}\def\l{\lambda}\def\n{\nu}\def\o{\omega}
\def\bfq{{\bf q}}\def\bfp{{\bf  p}}
\def\la{\langle}\def\ra{\rangle}\def\d{\delta}\def\bfr{{\bf r}}\def\k{\kappa}\def\G{\Gamma}\def\z{\zeta}
\def\calL{{\cal L}}\def\cL{{\cal L}}\def\GX{\Gamma_X}\def\o{\omega}\def\bfb{{\bf b}}
\def\BGs{\rm Bloom-Gilman\,duality}
\begin{document}



\title{NT@UW-20-09\\The Mystery of Bloom-Gilman Duality:  A   Light Front Holographic QCD Perspective}


\author{Aiden B.  Sheckler }

\author{Gerald A. Miller}
\email{miller@phys.washington.edu}
\affiliation{Department of Physics, University of Washington, Seattle, WA \ \ 98195, USA}

\date{\today}

\begin{abstract}
Light front wave functions motivated by holographic constructions are used to study $\BGs$, a feature of deep inelastic scattering.   Separate expressions for  structure functions  in terms of quark   and hadronic  degrees of freedom (involving transition form factors) are presented, with an ultimate goal of obtaining a relationship between the two expressions. A specific two-parton model is defined and resonance transition form factors are computed using  previously derived light front wave functions. A new  
 form of global duality (integral over all values of $x$ between 0 and 1) is derived from the valence quark-number sum rule.
Using a complete set of hadronic states is necessary for this new global duality to be achieved, and the  previous original work  does 
 not provide such a set. This feature is remedied by  amending the model to include a longitudinal confining  potential, and 
the resulting complete set is sufficient to carry out the study of  $\BGs$.
   Specific expressions for  transition form factors are obtained and all are shown to fall as $1/Q^2$, at asymptotically large values. This is because the Feynman mechanism dominates the asymptotic behavior of the model. These transition form factors are used to assess the validity of the  global  and local   duality sum rules, with the result that both are not satisfied within the  given model.
Evaluations of the hadronic expression  for $q(x,Q^2)$  provide more details about this lack. 
 This result is not a failure of the current model because it shows that   the observed validity of both global and local forms of duality for deep inelastic scattering must be related to a  feature of QCD that is deeper than completeness.
Our  simple present model  suggests a prediction that $\BGs$ would not be observed if deep inelastic scattering experiments were to be made on the pion.  The underlying origin of the duality phenomenon in deep inelastic scattering is deeply buried within the confinement aspects of QCD, and its origin  remains a mystery.
\end{abstract}




\maketitle
\section{Introduction}

Two  distinct facets of QCD are known. At asymptotically high-energies and momentum transfers $Q^2$ many hadronic observables can be computed using quarks and gluons as degrees of freedom and applying perturbation theory.   At low-momentum scales the theory is strongly coupled so that  hadronic degrees of freedom and  non-perturbative methods must be used. It has therefore been very  surprising that  examples in which the behavior of low-$Q^2$  cross sections can be  related, through suitable averaging procedures to those at high $Q^2$ actually exist. This phenomenon became known as quark-hadron duality. 
See the review~\cite{Melnitchouk:2005zr}.

Here we focus on deep-inelastic scattering from hadrons. Bloom and Gilman~\cite{Bloom:1970xb,Bloom:1971ye} studied data from the early DIS experiments at SLAC. They found that the inclusive structure function measured at low values of the hadronic final state mass, $W$, generally follows a curve that describes data at large values of $W$. In particular average values of the low $W$ cross sections were found to be  approximately equal to that of the high $W$ cross section, which approximately obeys Bjorken scaling. Moreover, averages of low $Q^2$ data over specific kinematic regions were found to be equal to the high $W$ cross section.

Bloom-Gilman duality was later examined in terms of an operator product expansion of moments of structure functions~\cite{DeRujula:1976ke,DeRujula:1976baf}. This work found a systematic  classification of terms responsible for duality, but did not describe how the physics of resonances transforms into the physics of scaling.  The subject of Bloom-Gilman duality was largely ignored for about 20 years.

The availability of high-luminosity beams of electrons at Jefferson Laboratory allowed the subject to be studied in great detail. A striking finding was that Bloom-Gilman duality appears to work at  values of $Q^2$ as low as about 1 GeV$^2$ or less~\cite{Niculescu:2000tk,Niculescu:2000tj,Ent:2000jj}.

Finding an elementary understanding of the origins of Bloom-Gilman duality has been elusive because it involves trying to build up structure function that is independent of $Q^2$ (except for the logarithmic corrections of QCD)  entirely out of resonances, each of which is described by a form factor that falls rapidly with $Q^2$.

The description of Bjorken scaling in deep inelastic (DIS) structure functions is most simply formulated in terms of the quark-parton model, which is well understood in terms of a twist expansion, but the physical final state is composed of hadrons. The validity of both approaches indicates that describing DIS in terms of hadronic degrees of freedom should be possible. One of the central mysteries of strong-interaction physics is how scattering from confined bound states of quarks and gluons can be consistent with Bjorken scaling, a property associated with free quarks.

Previous approaches to understanding the origins of $\BGs$, reviewed in~\cite{Melnitchouk:2005zr}, include QCD in 1 +1 dimensions~\cite{PhysRevD.14.3451}, phenomenological approaches~\cite{Domokos:1971ds,Domokos:1972yc}
harmonic oscillator models{~\cite{Isgur:2001bt,Jeschonnek:2003sb,Close:2002tm,PhysRevC.66.065205}  and other models ~\cite{Close:2001ha,Paris:2001pm,PhysRevC.65.035203}.  Non-relativistic potential models can describe or represent confining systems with an infinite number of bound states, thereby indicating  how it is that Bloom-Gilman scaling may arise, but are not properly relativistic. 
 This means that none 
obtain Lorentz invariant quark distributions that have the correct support properties of being non-zero only in the region where Bjorken $x$ varies between 0 and 1.  Understanding $\BGs$ requires models that are both confining and relativistic.

We aim   to understand $\BGs$ by using relativistic light-front wave functions obtained from
light front holographic QCD, an approach defined in the review~\cite{Brodsky:2014yha}, that provides a relativistic treatment of confined systems.  We briefly summarize following Ref.~\cite{Brodsky:2014yha}. 
Light-front quantization is a relativistic, frame independent approach   to describing  the constituent structure of hadrons. The simple structure of the light-front (LF) vacuum allows an unambiguous definition of the partonic content of a hadron in QCD and of hadronic light-front wave functions, the underlying link between large distance hadronic states and the constituent degrees of freedom at short distances.   The QCD light-front Hamiltonian $H_{LF}$  is constructed from the QCD Lagrangian using the  standard methods of quantum field theory~\cite{Brodsky:1997de}.  The spectrum and light-front wave functions of relativistic bound states are obtained from the eigenvalue equation  $H_{LF} \vert  \psi \rangle  = M^2 \vert  \psi \rangle$. It becomes an infinite set of coupled integral equations for the LF components  
 in a complete basis of non-interacting $n$-particle states, with an infinite number of components. This provides a quantum-mechanical probabilistic interpretation of the structure of hadronic states in terms of their constituents at the same light-front time  $x^+ = x^0 + x^3$, the time marked by the front of a light wave~\cite{Dirac:1949cp}.   The Hamiltonian eigenvalue equation  in the light front is frame independent.   The matrix diagonalization~\cite{Brodsky:1997de} of the LF Hamiltonian eigenvalue equation in four-dimensional space-time has not yet been achieved  because various technical and difficulties and because of a lack of understanding of the fundamental mechanism of confinement. Therefore other methods and approximations  are needed to truly  understand the nature of relativistic bound states in the strong-coupling regime of QCD.

To a first semiclassical approximation, where quantum loops and quark masses are not included, the relativistic bound-state equation for  light hadrons can be reduced to an effective LF Schr\"odinger equation. The technique is to  identify the invariant mass of the constituents  as a key dynamical variable. The invariant mass  measures  the off-shellness in  the LF kinetic energy,  so that it is the natural variable to characterize the hadronic wave function.  In conjugate position space, the relevant dynamical variable  is an invariant impact kinematical variable $\zeta$, which measures the separation of the partons within the hadron at equal light-front  time~\cite{deTeramond:2008ht}.  Thus,  the multi-parton problem  in QCD is reduced, in   a first semi-classical approximation, to an effective one dimensional quantum field theory by properly identifying the key dynamical variable.  As a result,  complexities of the strong interaction dynamics are hidden in an effective potential $U(\zeta)$,  but the central question -- how to derive the confining  potential from QCD, remains open.

It is remarkable that in the semiclassical approximation described above,  the light-front Hamiltonian  has a structure which matches exactly the eigenvalue equations in AdS space~\cite{Brodsky:2014yha}.  This offers the  possibility to explicitly connect  the AdS wave function $\Phi(z)$ to the internal constituent structure of hadrons. In fact, one can obtain the AdS wave equations by starting from the semiclassical approximation to light-front QCD in physical space-time.   This connection yields a  relation between the coordinate  $z$ of AdS space with the impact LF variable $\zeta$~\cite{deTeramond:2008ht},  thus giving  the holographic variable $z$ a precise definition and intuitive meaning in light-front QCD.

Light-front  holographic methods  were originally introduced~\cite{Brodsky:2006uqa,Brodsky:2007hb} by matching the  electromagnetic current matrix elements in AdS space~\cite{Polchinski:2002jw} with the corresponding expression  derived from light-front  quantization in physical space-time~\cite{Drell:1969km,West:1970av}. It was also shown that one obtains  identical holographic mapping using the matrix elements of the energy-momentum tensor~\cite{Brodsky:2008pf} by perturbing the AdS metric  around its static solution~\cite{Abidin:2008ku}, thus establishing a precise relation between wave functions in AdS space and the light-front wave functions describing the internal structure of hadrons.

The light front wave functions that arise out of this light front holographic approach provide a new way to study old problems that require the  use of relativistic-confining quark models. The study of $\BGs$ is an excellent example of such a problem. 

The present treatment is outlined next. Sect.~II is concerned with general definitions related to DIS for spin-less targets. Separate expressions for structure functions using   both quark and hadronic degrees of freedom are presented, with the fundamental aim of the paper to relate the two.  The global and local forms of $\BGs$ are reviewed briefly in Sect.~III. The next Section, IV,  is concerned with presenting the main features of light-front wave functions for two-parton systems that are obtained from the $z-\z$ connection. The valence quark number sum rule is discussed as a new global duality in Sect.~V. This relation is satisfied in models that provide a complete set of wave functions.
The existing light front wave functions, summarized in Ref.~\cite{Brodsky:2014yha}, are found to be incomplete in Sect.~VI. This is because excitations in the longitudinal degree of freedom are not incorporated. Completeness is implemented by including a longitudinal confining potential  in Sect.~VII. Evaluations of transition form factors are presented in Sec.~VIII.  The two-parton  model 
is shown to violate both the global and local forms of $\BGs$  in Secs.~IX and X.  The model's quark distributions are evaluated in Sec.~XI. The results are summarized and discussed in Sec.~XII.

\section{Deep Inelastic Scattering From a spin zero target    -General Preliminaries}
The hadronic tensor for a spin-less or spin-averaged target  is given by
\bea W^{\m\n}=(p^\m-{p\cdot  q\,q^\m\over q^2})(p^\n-{p\cdot q\,q^\n\over q^2}){W_2\over M^2}+ (g^{\m\n}-{q^\m q^\n\over q^2})W_1,\eea
with $p$ as the target  four-momentum ($p^2=M^2$) and $q$ that of the virtual photon.
The quantity $W^{\m\n}$  is the matrix element of a current-current correlation function:
\bea W^{\m\n}={1\over 4\pi}\int d^4\xi e^{iq\cdot\xi}\la p|j^\m(\xi)j^\n(0)|p\ra,\label{gen}\eea
with  normalization  $\la p'|p\ra= 2p^+ \d(p^+-{p'}^+) (2\pi)^3 \d(\bfp-\bfp')$.
 At large values of $Q^2$, the correlation is along the light cone.  For space-like four momenta $q^\m$ one may choose 
  the $z$ axis such that  $q^+=0$, for space-like values of $q^\m$, This enables the use of the  Drell-Yan formula~\cite{PhysRevLett.24.181} for form factors. With this notation  $q^3=-\nu$ , $|\bfq|=Q,\ \bfq^2 =Q^2, \,\vec {q}\,^2=\n^2+Q^2$.  Here bold-face indicates transverse vectors and $\vec q$ is the three-dimensional vector.
 
 \subsection{Quark Degrees of Freedom}
 
The expression \eq{gen} can be handled   by turning the product of currents into a commutator and then making the operator product expansion. The resulting leading-twist contribution to the quark distribution for a quark, $q(x)$ is given by
\bea q(x)=\int {dx^-\over 4\pi}e^{i x P^+x^-} \la p| \bar \psi(-x^-/2)\g^+\psi(x^-/2)|p\ra,\label{quark}\eea
where $\psi$ is a quark-field operator, the notation $x^-/2$ refers to $x^\m=(0,-x^-/2,{\bf 0})$,  and  $x=Q^2/2p\cdot q$.  Scale dependence due to QCD evolution is omitted in this paper. \eq{quark} is understood as involving wave functions evaluated at a given momentum scale.  Thus the structure functions discussed reflect the intrinsic bound-state structure of the hadrons, and thus  apply only at low resolution scales.
Bloom \& Gilman did not consider QCD evolution in their work. Accordingly such effect are not a part of the present initial modern analysis.

One extracts $W_2$ from $W^{\m\n}$ by using
\bea W_2=
 W^{++}.\eea 
 At leading twist
 \bea W_2=2M^2 {x\over p\cdot q}W_1,\label{rel}\eea
 where $M$ is the ground state mass. 

\subsection{Hadronic Degrees of Freedom}
 
An expression for $W^{++}$ can be obtaining using hadrons by 
 inserting a complete set of hadronic states $|X\ra$ between the current operators in \eq{gen}. 
 The use of $\la p|j^\m(\xi)|X\ra=e^{i (p-p_X)\cdot \xi}\la p|j^\m(0)|X\ra$, taking $\m=+,\n=+$  and integrating  over $\xi^+$ yields
\bea W^{++}=  \sum_X{1\over 2E_X} \d(p^-+q^--p_X^-)|\la X,x|j^+(0)|p\ra|^2,\label{wpp}\eea
in which it is understood that $\vec p_X=\vec p+\vec q=\vec q$, in the lab frame.  The notation $X,x$ is meant to include all degenerate states of the same angular momentum.
The factor of $2E_X$ in the denominator comes from the relativistic normalization of states.
The matrix elements of $j^+(0)$ are proportional to form factors:
\bea|F_{X,0}|^2\equiv \sum_{x}|\la X,x|{j^+(0)\over 2p^+} |p\ra|^2,\label{f0}\eea
with $p_X^+=p^++q^+=p^+$. The matrix element involves only internal coordinates of the wave functions,  and 0 denotes  the ground state in the lab frame with $\vec p=0$.

The argument of the delta function appearing in \eq{wpp} can be expressed in more detail
as \bea p^-+q^--p_X^-= M+\n-E_X(\vec q)= E_0+\n -\sqrt{M_X^2+Q^2+\n^2},\eea
where $M$ is the ground state energy
Then using $W^{++}=W_2$, we find
\bea &W_2=4M^2\sum_{X}{1\over 2\sqrt{M_{X}^2+Q^2+\n^2}} \d (M+\n -\sqrt{M_{X}^2+Q^2+\n^2}) |F_{X}|^2.\label{sum21}\\&
=4M^2\sum_{X}\d\bigg(M^2-M_X^2+Q^2({1\over x}-1)\bigg)|F_{X}|^2.\label{sum22}\eea

The study of Bloom-Gilman duality involves relating the quark \eq{quark} and hadronic expressions \eq{sum22}.
\section{Bloom-Gilman Duality}

Bloom and Gilman found that  the structure function in the resonance region, $W<2 $ GeV, was roughly equivalent on average to the scaling one, with averages obtained over the same region of the scaling variable:
\bea \o'={2M \nu +M^2\over Q^2}=1+{W^2\over Q^2} ={1\over x}+{M^2\over Q^2},\label{op}
\eea
where the invariant energy, $W$,
is given by
$ W^2=(p+q)^2=M^2 +2M\nu -Q^2,$
which is the square of the mass of a resonance that is excited. Application of \eq{op} requires that $\o'>0$ and $x>0$, so that
$Q^2>M^2$. Bloom \& Gilman noted that their sum rule is not valid if $Q^2$ is much less than 1 GeV$^2$.

Bloom and Gilman found that the data from the resonance region at low $Q^2$ oscillate around the scaling curve, with averages that are equal to the scaling curve. Furthermore the resonances move to  lower values of $\o'$ (higher values of $x$) with increasing $Q^2$. These observations were repeated at Jefferson Lab.

Bloom and Gilman quantified their studies  by observing the validity of a sum rule:
\bea {2M\over Q^2} \int_0^{\n_m}d\n\,\n W_2(\n,Q^2)= \int_1^{1+W_m^2/Q^2}d\o'\n W_2(\o'),\label{SR}\eea
where $W_2(\o')$ is scaling function obtained at large values of $Q^2$. The upper limit on the integration over  $\n$, $\n_m=(W_m^2-M^2+Q^2)/2M$ corresponds to the maximum value of $\o'=1+W_m^2/Q^2$, where $W_m\approx 2$ GeV.  The validity of this equation is known as {\it global duality}.

{\it Local duality} is said to exist if the equality of the averaged resonance and scaling functions holds over restricted regions of $W$. 
BG obtained an explicit expression by taking the difference between two versions of \eq{SR}  with different upper limits of integration.
\bea {2M\over Q^2} \int_{\n_a}^{\n_b}d\n\,\n W_2(\n,Q^2)= \int_{1+W_b^2/Q^2}^{1+W_a^2/Q^2}d\o'\n W_2(\o').\label{LSR}\eea
 
\section{Soft-wall Light Front Wave Functions in Light Front Holographic QCD}
Consider a  hadronic bound state of  two constituents, each of vanishing mass.  We follow the argument of~\cite{Brodsky:2014yha}.
The eigenmasses are given by
\bea& M^2 = \int_{0}^{1}\int\psi^*(x,\mathbf{k}  )\bigg[\dfrac{-\nabla^2_{\mathbf{k}  }  }{x(1-x)} \bigg]\psi(x,\mathbf{k}  )dx \ {d^2\mathbf{k}\over 16\pi^3}   + \text{interactions},\notag\\&
= \int_{0}^{1}\int\psi^*(x,\mathbf{b}  )\bigg[\dfrac{-\nabla^2_{\mathbf{b}  }  }{x(1-x)} \bigg]\psi(x,\mathbf{b}  )dx \ {d^2\mathbf{b}\over 4\pi}   + \text{interactions}\label{msq}
\eea
where
the relative coordinates are the transverse separation  $\bfb  $ and the momentum fraction $x$. 
 The invariant mass $M_{q\bar q}^2={\bfk^2\over x(1-x)}$. The canonically conjugate impact space variable is $\z^2=x(1-x)\bfb^2$. To a first approximation light front (LF) dynamics depend only on $M_{q\bar q}^2$ or $\z$, and the dynamical properties are encoded in the hadronic wave function $\phi(\z)$. Following standard procedures solutions in the product form,
\bea \psi(x,\z,\phi)=e^{i L_z\varphi}X(x){\phi(\z)\over \sqrt{2\pi \z}},\label{wf0}\eea
are sought. 
The quantity $L_z$ is the longitudinal component of the orbital angular momentum, an integer that can be positive of negative or 0.

One proceeds by writing the Laplacian operator in \eq{msq} in the polar coordinates $(\z,\phi)$;
\bea \nabla_\z^2={1\over \z}{d\over d\z}\big(\z{d\over d\z}\big)+{1\over\z^2}{\partial^2\over \partial\varphi^2}\eea
Then with the normalization
\bea \la X|X\ra=\int_0^1{dx\over x(1-x)}X^2(x)=1\eea one finds that \eq{msq} becomes
\bea 
M^2=\int d\z \phi^*(\z)\bigg(-{d^2\over d\z^2}-{1-4L^2\over 4\z^2}\bigg)\phi(\z)+\int d\z\phi^*(\z)U(\zeta)\phi(\z)\eea
in which an effective potential $U(\z)$ has been introduced to enforce confinement at some infrared scale, and $L$ is the absolute value of $L_z$.

The resulting  wave equation is 
	\bea 
	\bigg(-{d^2\over d\z^2}-{1-4L^2\over 4\z^2}+U(\z)\bigg)\phi(\z)=M^2\phi(\z)
	\label{we} \eea	
 The soft-wall model  ~\cite{Karch:2006pv}	
	\bea U(\zeta) = \kappa^4\zeta^2\label{sw}\eea
	where $ \kappa $ is the strength of the confinement, is used here. 
	This  form was used in ~\cite{PhysRevLett.102.081601}.  Later work~\cite{PhysRevD.91.045040,PhysRevD.91.085016}     introduced a constant term that depends on $L$, and relations between the baryon and meson spectrum was obtained. These results were derived from super-conformal algebra. In line with our goal of examining  Bloom-Gilman duality, the earlier form is used here to avoid a zero in the ground state mass. 
	
	The function $X(x)$ is to be determined. A popular choice has been to simply use $X(x)=\sqrt{x(1-x)}.$  This is followed here. We'll show below that this choice does not yield a complete set of wave functions in three-dimensional space.

		The associated eigenvalues are :
	\bea M^2_{nL} = \kappa^2(4n+2L+2) . \label{mass1}\eea
The  light-front wave function (\eq{wf0}) is obtained  by  	solving \eq{we} with the soft-wall potential of \eq{sw}, and using 
$\z^2=x(1-x)b^2$ with the result:
	\bea \psi_{nL_z}(\bfb,x) = A_{nL} \sqrt{x(1-x)}{e^{i L_z \varphi} \over\sqrt{2\pi}}e^{-\k^2b^2 x(1-x)/2}(\sqrt{x(1-x)}b)^{L} L_n^{L}(\k^2 b^2 x(1-x)).\label{wf}\eea
		where  $ b $ is the transverse radius, $ \varphi $ is the transverse angle, and $ x $ is the plus-momentum ratio of one of the quarks.  
The factor normalizing the wave function to unity is:
%
$ A_{nL}= {\sqrt{2} }\sqrt{n!\over (n+L)!}.$
The spin-dependence is taken be a simple delta function, setting the helicity of the anti-quark to the negative of the quark~\cite{Lepage:1980fj,Brodsky:2007hb}. The action of the $\g^+$ operator in $j^+$ conserves the spin of the struck quark so that all of the states, $X$ entering in \eq{sum22} have the same spin wave function.

The next step is to evaluate the  transition form factors.
 For a quark-anti-quark system, of unit charge, with the given space and spin dependence,  in the Drell-Yan frame,
these are given by the expression:
		\bea &F_{X0}(\bfq) \equiv F_{nL_z} (\bfq) =\int_{0}^{1} dx \int d^2\bfb  \ \ \ e^{i {\bfq} \cdot {\bfb} (1-x)} \ \psi^*_{nL_z}(x,{\bfb}) \ \psi_{00}(x,\bfb),\label{tf} \eea
with the notation $ (\bfq\cdot \bfq=Q^2)$. This form factor is the same as would be obtained if only a single quark of unit charge interacts electromagnetically.
		
\section{Valence Quark number sum Rule- a New Global Duality}\label{sec:sum}
In this two-parton model the quark and anti-quark have the same relative wave function and therefore the same quark distribution.
The form factor can be obtained as if only a single quark of unit charge interacts electromagnetically. In this case the form factors of \eq{f0} contribute to the single flavor $q(x,Q^2)$.
This means that the sum over $X$ in \eq{sum22} contributes only to $q(x)$, and 
we may take
   the quark distribution $q(x)$ as given by
\bea q(x)=W_1={p\cdot q\over 2M^2 x}W_2={\nu\over 2Mx}W_2.\label{qW}\eea\\


An interesting sum rule may be derived from the valence  quark number sum rule:
\bea \int_0^1q(x)dx=1.
\eea
The integral over $x$ can be evaluated from the hadronic degrees of freedom, using  \eq{qW} and \eq{sum22} we examine the quark number sum rule 
 to find
\bea q(x,Q^2)={Q^2\over x^2} \sum _{X}\d\bigg(M^2-M_X^2+Q^2({1\over x}-1)\bigg)|F_{X}|^2 \label{qx}\eea
Noting that  $F_X$ depends on $Q^2$, not on $x$, the integral over all values of $x$ yields
\bea& \int_0^1q(x)dx=\sum_X|F_{X}|^2.\label{ngd}\eea
Thus the quark number  sum rule is satisfied if and only if the completeness relation
\bea \sum_X|F_{X}|^2=1\label{sum23}\eea is satisfied. 

This result  \eq{ngd}  amounts to a new form of global duality: At any given value of  $Q^2$ the sum of the squares of all of the form factors, computed using hadron degrees of freedom, satisfies the quark number sum rule. The derivation presented here   uses  a single quark flavor and unit quark charge. The same global duality may be  obtained in terms of the usual more general parton model conditions. It is a consequence of baryon number conservation.\\

\section{Lack of Completeness}
In the current model the sum appearing  in \eq{sum23} is given by
\bea S(Q)=\sum_{n=0}^\infty\sum_{L_z=-\infty}^\infty \bigg|F_{nL_z}(Q)\bigg|^2,\label{complete}\eea
Satisfying the  sum rule requires $S(Q)=1$.

Let's evaluate this quantity. 
    To see this more explicitly consider
  \bea& S(Q)\equiv \sum_{n=0}^\infty\sum_{L_z=-\infty}^\infty  \bigg|F_{nL_z}(Q) \bigg|^2 \\&=\int dx\int d^2b e^{i\bfq\cdot\bfb(1-x)}\psi_{n,L_z}^*(x,\bfb)\psi_{00}(x,\bfb) 
 \int dx'\int d^2r' e^{-i\bfq\cdot\bfb'(1-x')}\psi_{n,L_z}(x',\bfb')\psi_{00}(x',\bfb') 
\eea
  Change variables to $\bfzeta =\bfb\sqrt{x(1-x)}$. Then  use $d^2b x(1-x)= d^2\z$ and define
  \bea \phi_{nL_z}(\bfzeta)\equiv {\psi_{nL_z}(x,\bfzeta)\over \sqrt{x(1-x)}}.\eea
  The $\phi_{nL_z}(\bfzeta)$ are standard two-dimensional harmonic oscillator wave functions.
   Then
  \bea &S(Q)=\sum_{n=0}^\infty\sum_{L_z=-\infty}^\infty \int dx\int d^2\zeta e^{i\bfq\cdot\bfzeta\sqrt{1-x\over x}}\phi_{n,L_z}^*(\bfzeta)\phi_{00}(x,\bfzeta) 
 \int dx'\int d^2b' e^{-i\bfq\cdot\bfzeta'\sqrt{1-x'\over x'}}\phi_{n,L_z}(\bfzeta')\phi_{00}(\bfzeta') 
\eea
But  standard 2D ho wfs obey
\bea \sum_{n=0}^\infty\sum_{L_z=-\infty}^\infty \phi_{n,L_z}^*(\bfzeta) \phi_{n,L_z}(\bfzeta')=\d(\bfzeta-\bfzeta'),\eea
so that
  \bea &S(Q) = \int dx\int dx' \int d^2\zeta e^{i\bfq\cdot\bfzeta(\sqrt{1-x\over x}-\sqrt{1-x'\over x'})} |\phi_{00}(\bfzeta)|^2\eea
The integral over $\bfzeta$ can be done so that
\bea S(Q)=\int dx\int dx' e^{- 1/4 Q^2 (\sqrt{1-x\over x}-\sqrt{1-x'\over x'})^2}\label{SQ}\eea
The validity of the sum rule requires that $S(Q)=1 $ for all values of $Q$. This is true only for $Q=0$. For all other values, it is manifest that $S(Q>0)<1$. Completeness is not satisfied.
Numerical work shows that $\lim_{Q\to \infty}S(Q)\sim {1\over Q },$
and  one can analytically show $\lim_{Q\to \infty}S(Q)\approx\frac{\pi ^{3/2}}{8 Q}$  by using the method of steepest descent. The net result is that 
 instead of unity one gets 0 for large enough values of $Q$.

This study of $S(Q)$ shows that 
 \eq{complete} is NOT satisfied in the current model. This is because the wave function given above  in \eq{wf}  are complete only in the two-dimensional  $\bfzeta$ space, not in the $x$ space.
\section{Implementing completeness}

The functions of \eq{wf} do not form a complete set over the three-dimensional space space $x,\bfr$ because $X(x)=\sqrt{x(1-x)}$ is not a complete set of wave functions in $x$ space.  The need to include excitations of longitudinal modes has  noted in 
Ref.~\cite{Li:2015zda}.

Here we construct 
a set of wave functions in $x$ space in which the given $X(x) $ corresponds to a ground state. This is done by generalizing the interaction to include a longitudinal potential 
\bea U_\calL(x)=-\l \k^2 {d\over dx}x(1-x) {d\over dx}.\label{UL}\eea
This  interaction   is approximately harmonic oscillator potential in the longitudinal variable $\tilde z$ of \cite{Miller:2019ysh}, with $\tilde z^2=-{\partial^2 \over \partial x^2}$. The factor $x(1-x)$ appears here and in the transverse soft-wall potential of \eq{sw} when it is expressed in terms of light-front variables.

This added potential  gives a contribution to the square of the mass, $M_\calL^2$ given by
\bea M^2_\calL=-
\lambda\k^2\int_0^1dx {X(x)\over \sqrt{x(1-x)}}{d\over dx}x(1-x) {d\over dx} {X(x)\over \sqrt{x(1-x)}},\label{L}\eea
where $\l$ is a dimensionless number.
The normalized  solutions to the related differential equation are the functions 
\bea  {X(x)\over \sqrt{x(1-x)}}=\sqrt{2\calL+1} P_\calL(2x-1),\label{UL1}\eea 
where $P_\calL$ is a Legendre polynomial (with $\calL$ an integer) with
eigenvalues 
$M_\calL^2=\lambda\k^2\calL(\calL+1)$.
For $\calL=0,\, X(x)=\sqrt{x(1-x)}$ so previous results for the spectrum are preserved and correspond to modes with $\calL=0$.\\

The value of $\l$ should be determined by an appropriate symmetry. We hope that the need to satisfy completeness will lead to future work on this topic and leave 
finding such a symmetry for future work and future workers.  Our only purpose here is to study Bloom-Gilman duality. A complete set of states is needed to do that, as shown in Sect.~\ref{sec:sum}.
 Here we note that setting $\l=2$ means the low-energy part of the spectrum obtained in Refs.~\cite{PhysRevD.91.045040,PhysRevD.91.085016}  is not changed. 

The net result is that   light-front wave functions that provide the necessary complete set are given by
\bea\Psi_{nL_z\calL}(x, \varphi, \bfr) = \psi_{nL_z}(\bfr,x) \sqrt{x(1-x)} P_\calL(2x-1)\eea
  Using $\Psi_{n,L_z\calL}$ satisfies the completeness relation  in the $(x,\bfb)$ space because the Legendre polynomials form a complete set of orthogonal polynomials.
 

The spectrum is now given by
\bea
M^2_{nL\cL}(\l)=\k^2(4n +2L+ 2\calL(\calL+1)+2).\label{spect}\eea


\section{Transition  Form Factors}
The transition form factors must be evaluated in preparation for calculating $W_2$ as given by the hadronic expression of \eq{sum22}.
Specific expressions for  these form factors have not been presented previously. 
\\

Let's start with the $\calL=0$ sector. The angular integral appearing in $d^2b$ can be done in closed form with the result:
		\bea F_{nL_z}(\bfq)
= (-i)^{L_z}  {2  } \sqrt{n!\over (n+L)!}\int_0^1dx\int_0^\infty dz\,  z^{L+1}e^{-z^2}L_n^L(z^2)J_L\big(Qz\sqrt{1-x\over x}\big)\eea
Evaluation leads to
	\bea F_{nL_z}(Q) =   {(-i)^{L_z}}\sqrt{\frac{1}{n!(n+L)!}} \ \int_{0}^{1}dx \ \bigg( \frac{Q}{2\kappa}\sqrt{\frac{1-x}{x}} \bigg)^N  e^{-\frac{Q^2}{4\kappa^2}\frac{1-x}{x}},\label{fint} \eea
in which $|\bfq|=Q$ is used and $N\equiv 2n +L$. 
	
	This integral is found analytically in terms of an incomplete Gamma function as 
	\bea  F_{nL_z}(Q) =(- i)^{L_z} \ \sqrt{\frac{1}{n!(n+L)!}}\frac{2^{-(N+1)}}{N}\Gamma\Big( \frac{N+2}{2} \Big)\Bigg( -2^{N}\frac{Q^2}{\kappa^2} \ + \  e^{Q^2/4\kappa^2}\Big(\frac{Q}{\kappa}\Big)^{N}\Big( \frac{q^2}{\kappa^2} + 2N \Big) \ \Gamma\Big(\frac{2-N}{2}, \frac{Q^2}{4\kappa^2}\Big)\Bigg)\label{ff} \eea

An alternate evaluation of the integral  that allows the asymptotic limit of the transition form factors to be obtained is given next.


 
 Define \bea I_N\equiv  \int_{0}^{1}dx \ \bigg( \frac{Q}{2\kappa}\sqrt{\frac{1-x}{x}} \bigg)^N  e^{-\frac{Q^2}{4\kappa^2}\frac{1-x}{x}}\eea
with \bea F_{nL_z}(Q)=(- i)^{L_z} \ \sqrt{\frac{1}{n!(n+L)!}}I_N(Q).\label{FNL}\eea
 Let $K\equiv N/2$, $z\equiv  {Q^2\over 4\k^2}$ and $u\equiv {1-x\over x}$. Then
 \bea I_K(z)=z^K \int_0^\infty {du\over (1+u)^2}u^K e^{-z u}=z \Gamma (K) \left(e^z (K+z) E_K(z)-1\right), K>0,\eea
 with
 \bea E_K(z)=\int_1^\infty {e^{-zt}\over t^K}dt,\eea
 and
 \bea I_0(z)= \int_0^\infty {du\over (1+u)^2} e^{-z u}=1-z  e^z \int_{z}^\infty{e^{-t}\over t}dt\label{i0}\eea

It is useful to consider the asymptotic values of the transition form factors for fixed values    of $K$ and $z\gg K$. This is because Bloom \& Gilman found that one of the requirements for local duality is that all of the transition form factors have the same dependence on $Q^2$.
 First note that for integer values of $K$ one may write 
\bea I_K(z)=z^K (-{\partial\over \partial z})^K I_0(z).\label{iter}\eea

The asymptotic limit of $z\to\infty$ is obtained by integration by parts of the first of the expressions for $I_0(z)$:
\bea  I_0(z)={-1\over z}\int_0^\infty {du\over (1+u)^2} {d\over du}e^{-z u}={1\over z} +{\cal O }({1\over z^2}+\cdots)\eea
Keeping the leading term and carrying out the derivatives of \eq{iter} gives
\bea I_K(z) \sim {K!\over z}={K\G(K)\over z},\label{limx}\eea which leads to the  same result as taking the limit of \eq{ff}. The latter expression works for half-integer values of $L$.

The net result is that as $Q^2$ approaches $\infty$ for $z=Q^2/4\k^2\gg K.$
\bea \lim_{Q^2\to\infty}F_{nL_z}(Q^2) = {4(n+{L\over2})!\over Q^2}i^{L_z}{1\over \sqrt{n!(n+L)!}}
\label{fa}\eea\\


\noindent This limit is not to be used in evaluating the form factors necessary to compute $q(x)$  because the assumption that $z\gg K$  is violated in doing the sum over states which goes in principle to infinite values of $K$.   
\\

The universal $1/Q^2$ behavior shown in \eq{fa} occurs because the integral of \eq{i0} is dominated by small values of $u$ which corresponds to large values of $x$. This model presents an example of the Feynman model~\cite{Feynman:1973xc} of form factors in which the dominant contributions to the form factors occur when one quark, carrying nearly all of the momentum of the hadron is turned around by the virtual photon.

 \subsection{$\calL>0$}
 
 We now need 
 \bea &I_{K,\calL}(z)\equiv z^L \int_0^\infty {du\over (1+u)^2}u^K e^{-z u} P_\calL({1-u\over 1+u})\label{newI} \eea
Using the same arguments as before, we can show that
\bea \lim_{z\to\infty}I_{K,\calL}(z)={K\G(K)\over z}.\eea
This is because the Legendre polynomial in \eq{newI} is unity at $u=0$.

We have not been able to obtain a closed form expression for a general value of $\calL$. Instead we evaluate term by term
\bea &I_{K,1}(z)=-z \Gamma (K) \left(e^z \left((K+z)^2+z\right) E_K(z)-K-z-1\right),\,K>0)\\&
I_{0,1}(z)=z \left(e^z (z+1) \int_z^\infty dt {e^{-t}\over t}-1\right)
\eea

\bea &I_{K,2}(z)=  z \Gamma (K) \left(e^z \left((K+z) \left((K+z)^2+3
   z\right)+z\right) E_K(z)-K^2-K(2 z+1)-z (z+3)-1\right),\,K>0\\&
   I_{0,2}(z)=z (2 + z -e^z (1 + z (3 + z))  \int_z^\infty dt {e^{-t}\over t})
   \eea
    
    Similar expressions can be obtained for any value of $\cL$. The net result is that
    \bea F_{nL_z,\cL}(Q)=(- i)^{L_z} \ \sqrt{\frac{1}{n!(n+L)!}}\sqrt{2\cL+1}I_{K,\cL}(Q),\label{fff}\eea
    with $K=n+L/2$.
    
    \section{Bloom-Gilman global duality is not satisfied by this model}
       \begin{figure}[h] \label{duality1}
		\includegraphics[width=.4\textwidth]{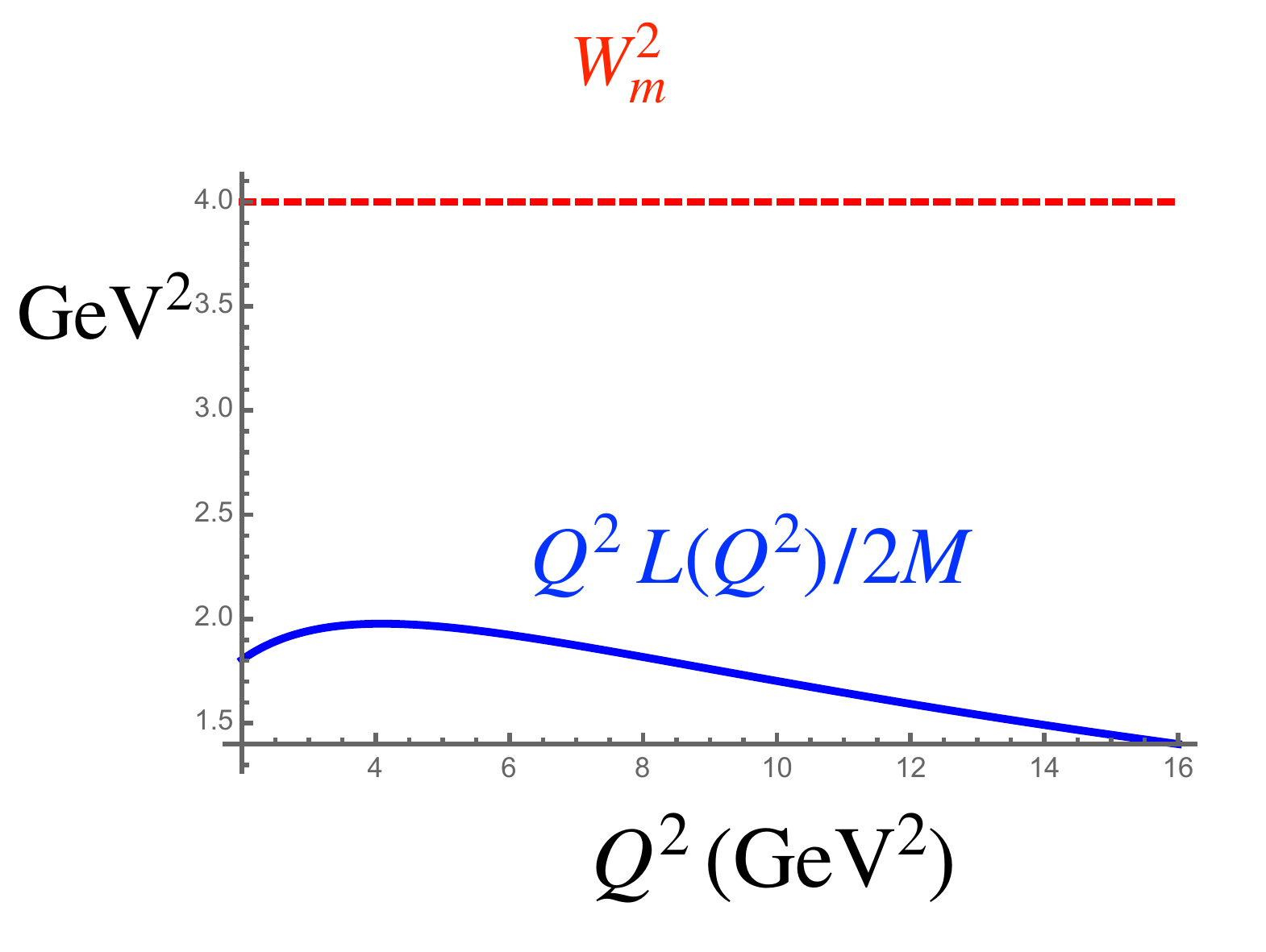} 
     \caption{$Q^2 L(Q^2)/2M\, {\it vs.}\, W_m^2$ 
     }\end{figure}

    Prior to evaluating $q(x)$ it is useful and possible to see if the sum rule of \eq{SR} is satisfied. 
    The right hand side of that equation is to be obtained by the high $Q^2$ expression for $q(x)$. 
    This quantity is obtained from  \eq{quark} by using the field expansion, in the standard expressions of the two-parton Fock-space component wave function. The result using the model of Sect.~IV is that
    \bea q(x)=\int d^2b \big|\psi_{00}(b,x)\big|^2 =1.\label{model}\eea 
This expression involves an integral over all values of $\bfb$ and therefore is equivalent to an integral over all values of $\bfk$. This latter integral corresponds to an infinite momentum transfer scale~\cite{Lepage:1980fj}.  Thus QCD evolution of the distributions is not discussed further. The $Q^2$ dependence of  $q(x,Q^2)$ of \eq{qx} arises from the  $Q^2$ dependence of the resonance form factors. \\

The result $q(x)=1$ requires further comment because it displays the unrealistic nature of the present model. According to quark counting rules, see e.g.~\cite{Brodsky:1994kg}, the structure function of a  two-parton state should fall as $1-x$ for large values of $x$, with QCD-DGLAP evolution providing a faster fall-off.  Bloom and Gilman~\cite{Bloom:1970xb,Bloom:1971ye} use quark counting rules in the form of the Drell-Yan-West~\cite{Drell:1969km,West:1970av} relations to understand their duality. As shown by Drell and Yan~\cite{Drell:1969km},  a quark structure function  varies at large $x$ as  $ ( 1 - x ) ^ {2 n - 1}$  if  the corresponding form factor  $F(Q^2)\sim (1/Q^2)^n$ at large values of $Q^2$. The present model violates this relation because, as shown above, all of the form factors have the same asymptotic $1/Q^2$ behavior, but the structure function is constant.  In the present model, the asymptotic nature of the form factor is determined by the Feynman mechanism. The quark counting rules are based on vector meson exchange between the quarks, so that the important region occurs when all of the partons are on top of each other. Feynman~\cite{Feynman:1973xc} argued  that is not realistic.\\

The validity of the Feynman {\it vs} quark-counting picture is one of the important issues related to color transparency~\cite{Frankfurt:1994hf}, the interesting quantum invisibility predicted to occur as the result of vanishing of initial or final-state interactions.  Color transparency cannot 
 occur    if the Feynman mechanism is dominant~\cite{Frankfurt:1993es}.\\

    The purpose of the present paper is to investigate the stated model. Although simple it has the correct completeness and support properties that
    enable study of Bloom-Gilman duality. Therefore we investigate the sum rule \eq{SR} by 
    using $q(x)=1$. This   allows the right-hand-side $\equiv R(Q^2)$ of \eq{SR} to be  evaluated immediately:
    \bea  R(Q^2)\equiv \int_1^{1+W_m^2/Q^2}d\o'\n W_2(\o')=2M \int_1^{1+W_m^2/Q^2}d\o' x 
    ,\eea 
    with $x=1/(\o'-M^2/Q^2)$.
    Evaluation yields 
    \bea R(Q^2)=2M\ln{Q^2+W_m^2-M^2\over Q^2-M^2},\eea
    and
    \bea \lim_{Q^2\to\infty} R(Q^2) =2M{W_m^2\over Q^2}\eea 

This result of using the stated model is already in violation of the Bloom-Gilman condition that it be independent of $Q^2$. The independence could be obtained by taking $Q^2$ to infinity, so that $R\to0$, and the violation of global duality is assured.

Next examine the left-hand side of \eq{SR} ($\equiv L(Q^2)$). First  that the lower limit on the integral over $\n$ is determined by the largest value of $x\,(=1)$ for which  $\n W_2$ is non-zero. This is given by $Q^2/(2M)$. Converting the integral over $\n$ to one over $x$ gives
\bea L(Q^2)={2M} \int^1_{Q^2\over (Q^2+W_m^2-M^2)}      {dx\over x} q(x,Q^2)\label{L}\eea
in which $q(x,Q^2)$ is   to be obtained from the hadronic expression,  \eq{sum22} . Simplifying  the argument of the delta function leads to:
\bea q(x,Q^2)=\sum_X |F_X(Q)|^2 \d(x- x_{X}),\eea
where $X\equiv (n,L_z,\cL)$ and 
$ x_X={Q^2\over Q^2+ 2\k^2( 2n +L+\cL(\cL+1))}.$
 
 Thus
 \bea L(Q^2)=2M\sum_{n,L_z,\cL}{Q^2+ 2\k^2( 2n +L+\cL(\cL+1))\over Q^2} F^2_{nL_z\cL}(Q^2)\Theta(W_m^2 -2\k^2(2n+L+\cL(\cL+1))\label{lsum}\eea
 If $R$ were equal to  $L$ the sum appearing in \eq{lsum} would need to vary as $1/Q^2$. This seems unlikely because for finite values of $M_X^2$, each form factor varies as $1/Q^2$, leading to an overall dependence varying approximately as $1/Q^4$. The asymptotic limit is not precisely accurate, but nevertheless $L(Q^2)$ falls much faster than $1/Q^2$. 
 
 This dependence is shown in Fig.~1. To obtain this figure we take $M^2=2\k^2$, with $M$ the nucleon mass and $W_m =2$ GeV, so that $2n+L+ 2\cL(\cL+1)\le4$. The result is shown in Fig.~1. Global duality would hold if $Q^2R(Q^2)/2M=Q^2L(Q^2)/2M$ {\it i.e,} if the blue and red curves were equal.
\section{Bloom Gilman Local Duality is not satisfied by the model}
Local duality is studied through \eq{LSR}. If the upper and lower limits are taken to encompass one resonance, $X$, this equation leads to the relation
\bea {1\over  x_X}F^2_{X}=\ln{W_+^2+Q^2-M^2\over W_-^2+Q^2-M^2},\label{lsr}\eea
with $W_\pm= M_X\pm  \g/2$. Here the value of $\g$ can be any energy less than the minimum spacing between levels, $2\k^2$.
A first glance indicates that the relation \eq{lsr} is not generally satisfied because the right hand side depends on $\g$, but the left hand side does not. Moreover, the right-hand side  falls as $1/Q^2$ for large values of $Q^2$, but the left-hand side falls as $1/Q^4$.

An explicit calculation is made by choosing  the lowest energy resonance with $n=0,\,L=1.\,\,\cL=0$ as an example. The results, shown in Fig.~2 are that indeed the local duality relation is not satisfied.

   \begin{figure}[h] \label{duality2}

		\includegraphics[width=.4\textwidth]{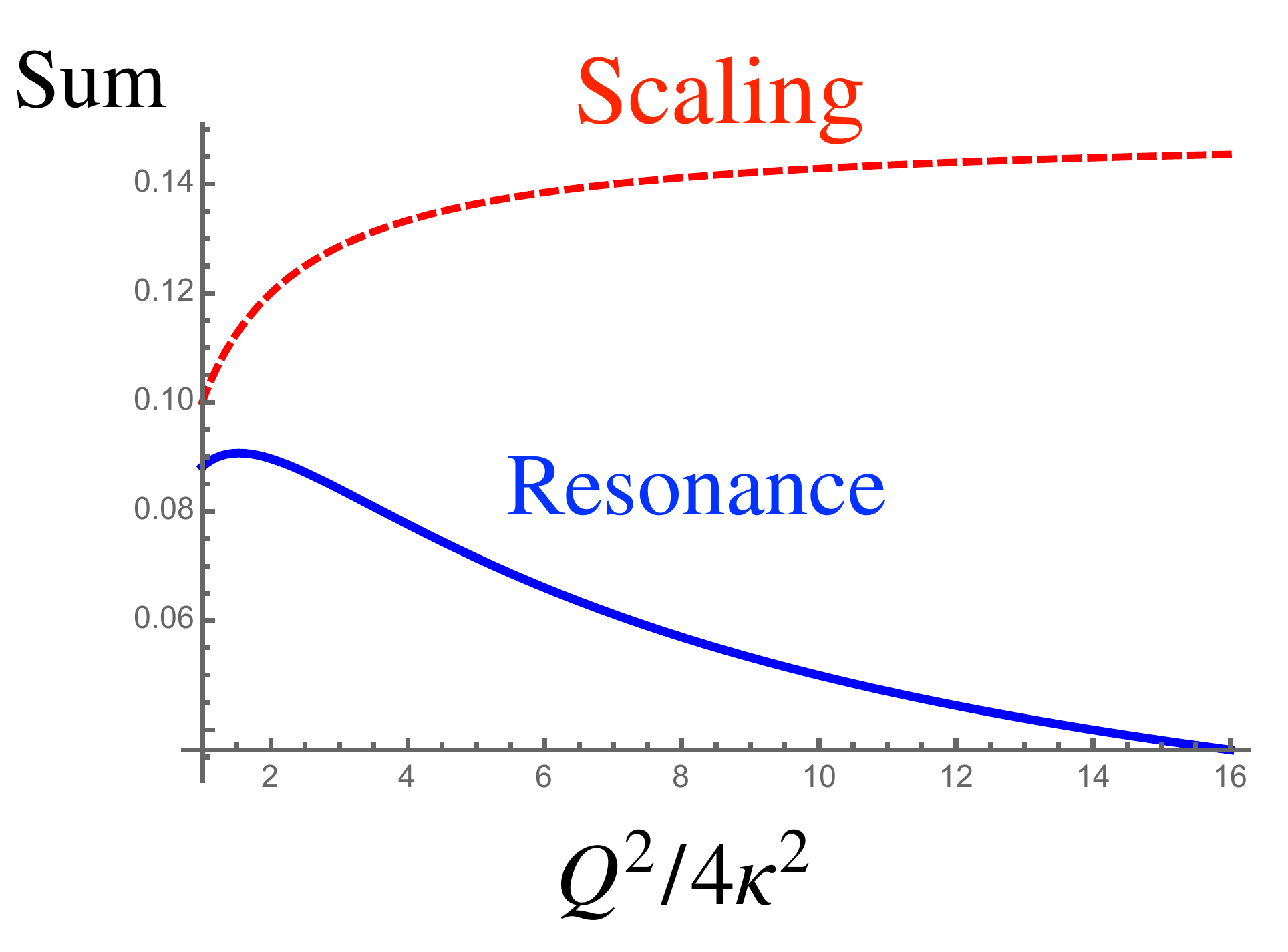} 
     \caption{Left- (blue, dashed,Resonance) and right- (red,solid,Scaling)  hand sides of \eq{lsr} times $Q^2/4\k^2$, with $\g=0.125\k$.} \end{figure}  

 \section{Evaluate $q(x,Q^2)$}
\begin{figure}[h] \label{duality3}

		\includegraphics[width=.34\textwidth]{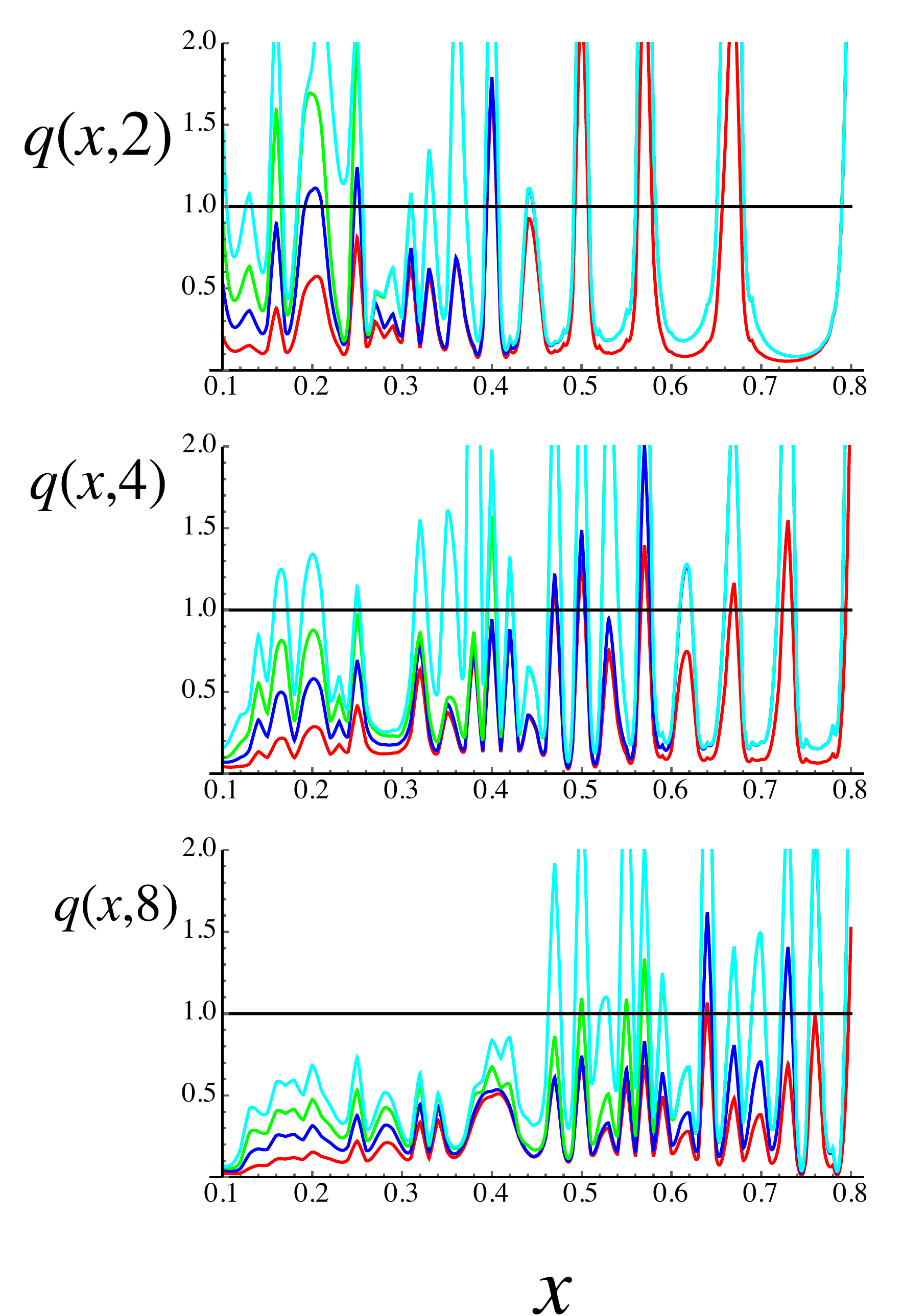} 
     \caption{$q(x,z)$ for $z=Q^2/4\k^2\,=\,2,\,4,\,8$. The    four curves in each figure are $q_{\cL=0}(x),\, q_{\cL=0}(x)+q_{\cL=1}(x),
    q_{\cL=0}(x)+q_{\cL=1}(x)+q_{\cL=2}(x) ,\,q_{\cL=0}(x)+q_{\cL=1}(x)+q_{\cL=2}(x) +q_{\cL=3}(x).$} \end{figure}  

\begin{figure}[h] \label{duality4}
		\includegraphics[width=.34\textwidth]{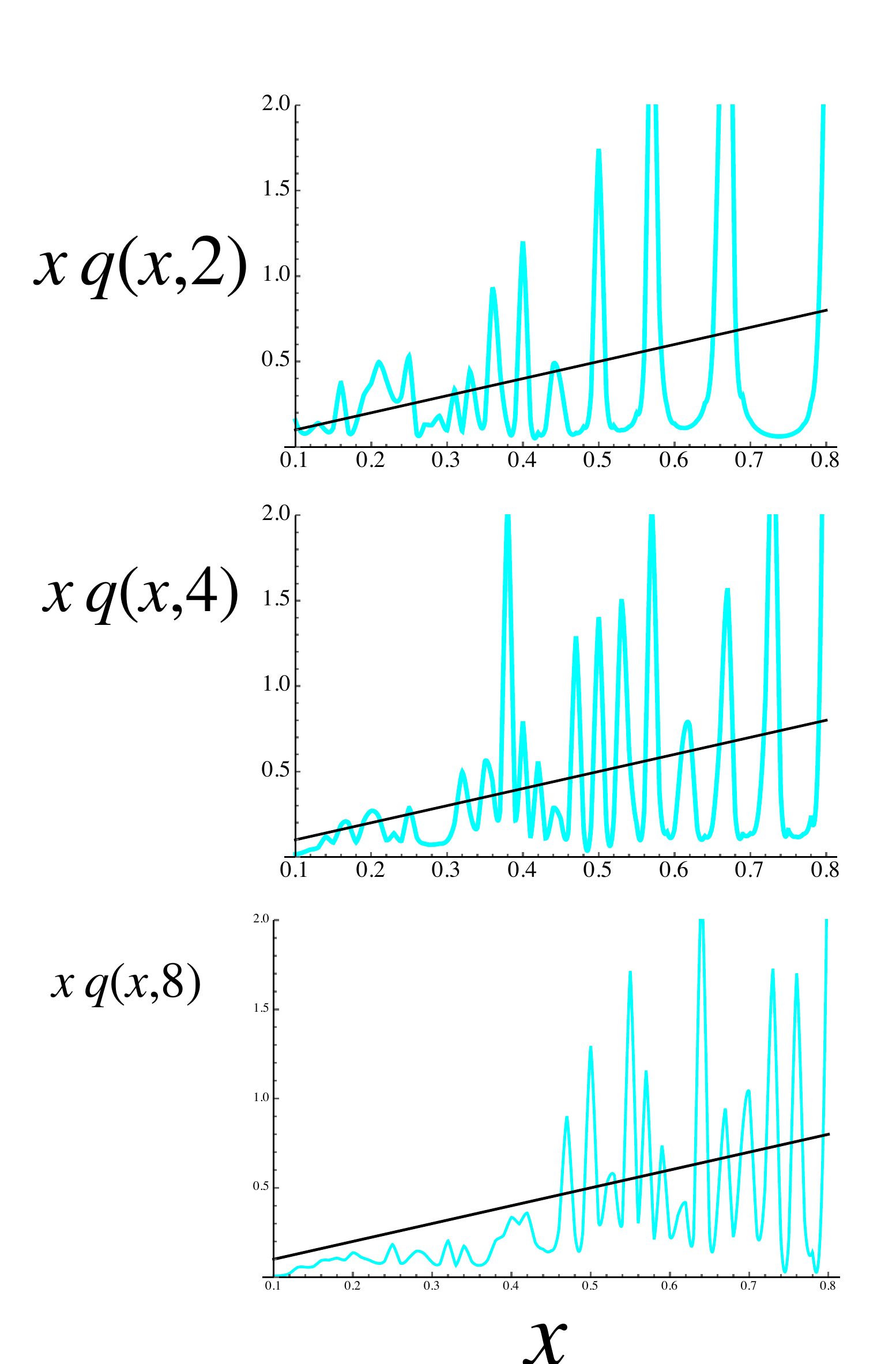} 
     \caption{$x(q_{\cL=0}(x,z)+q_{\cL=1}(x)+q_{\cL=2}(x) +q_{\cL=3}(x))$ for $z=Q^2/4\k^2\,=\,2,\,4,\,8$. The lines are drawn to guide the eye.} \end{figure}  

The quantity to evaluate is $q(x,Q^2)$ of \eq{qx}.  The need to include a non-zero width of the excited states in studying duality  has been noted  in~\cite{Isgur:2001bt,Jeschonnek:2003sb}.  Moreover, excited states do have non-zero widths. This is addressed next using a Breit-Wigner form.






The starting point is the function $\d(M^2-M_X^2+Q^2(1/x-1) =\d(M_X^2 -W^2),$ where $W^2=(p+q)^2$.
The value of $x$ that yields a vanishing argument is defined to be $x_X$ with
\bea x_X(Q^2)={Q^2\over M_X^2-M^2+ Q^2},\label{xx}\eea
showing that the contribution of a given resonance moves to larger values of $x$ as the value of $Q^2$ increases.

 It is useful to relate $\delta(W^2-M_X^2)$  to a delta function  $\delta(x-x_X)$, allowing a direct study of the new global duality of
 Sect.~V.  This may be done by first  using $ \d(M_X^2 -W^2)= {1\over 2W}\d(M_X-W)={1\over 2W} \lim_{\G\to0}{1\over \pi}{\G/2\over W-M_X+i\G/2}$. Then apply a  small-width approximation, $W-M_X =\sqrt{Q^2({1\over x}-1)+M^2}-\sqrt{{Q^2({1\over x_X}-1)+M^2}}\approx- (x-\xX) {Q^2\over 2M_Xx_X^2} $ so that \bea&
\d(M_X^2 -W^2)\to{\GX \over \pi } {M_Xx_X^4\over Q^4(x-\xX)^2+\GX^2x_X^4M_X^2)} \equiv \hat\d(W^2-M_X^2),
\label{final}\eea
The final step is to make the appearance of $\delta(x-x_X)$ explicit:
\bea  &\hat\d(W^2-M_X^2)= {1\over\pi}{\ve_X \over(x-x_X)^2+\ve_X^2} {x_X^2\over Q^2}
=\hat\d(x-x_X){x_X^2\over Q^2} \label{delta}\eea
with
$ \ve_X\equiv {\G_X M_X x_X^2\over Q^2}$.
Then \bea q(x,Q^2)=\sum_X\,{x_X^2\over x^2}\big|F_X(Q^2)\big|f\hat\d(x-x_X)\equiv \sum_{\cL=0}^3 q_\cL(x,Q^2).\label{final}\eea
The factor $f$ is insures that $\int_0^1 dx\hat\d(x-x_X)=1$, and  $f$ approaches unity as $\ve_X$ approaches 0. The second form shows the sum of states with different values of $\cL$. Each $q_\cL(x,Q^2)$ includes a sum over values of $n$ and $L_z$. A maximum value of $\cL=3$ is used. Although, not complete, this is sufficient to display the main points, and is necessary because an infinite number of values is needed for completeness.

The next step is evaluate  \eq{final}. Note that the sum over $X$ involves summing over $n,L_z=\pm L$ and $\cL$, including all states
The results of evaluating \eq{final}  are shown for $z=Q^2/(4\k)=2,4,8$ in Fig.3.  This corresponds to $Q^2 =8,16,32 \k^2=4,8,16 \,\rm M^2$. The width is taken as $\G/\k=0.05=0.2M/\sqrt{2}$. This is a constant width of  space 66 MeV, with  $M$ as the nucleon mass.
In the spirit of the model, the width is taken to be small~\cite{Isgur:2001bt,Jeschonnek:2003sb}. The maximum  number of states in the sum \eq{final} is increased until convergence is reached. 
In each figure the lowest curve is obtained using $\cL=0$, then the effects of $\cL=1$ are added leading to the next lowest curve. Then the effects of adding states with $\cL=2$ and $\cL=3$ are included and result in two more curves. 

Focusing first on $z=2$ and examining  the lowest curve, one sees a spiky structure due to the resonance curves. Decreasing the value of the width leads to similar results, with narrower widths and higher peaks. The same pattern is obtained when
 adding the effects of states with $\cL>0$, with different values of $M_X$, \eq{spect}, leading to different peaks that increase the distribution function at low values of $x$, as expected from \eq{xx}. 
The  contributions  of  states with $\cL>0$ don't contribute  at higher values of $x$ due to the delta function appearing in \eq{final}.
The solid line in each panel shows the scaling result $q(x)=1$. The resonance curves oscillate about the scaling curve for the larger values of $x$, but not for the lower ones.

 The patterns for different values of $\cL$ are  also seen as the value of $z$ increases. A detailed difference  the effect at high values of $x$ are larger. Another is that the curves fall further below the scaling result, and  the general tendency of the magnitude of $q(x,Q^2)$ to decrease as $Q^2$ increases is seen. This is another example of the model's failure to achieve local duality.

The  Bloom \& Gilman work found duality in $\nu W_2$ which is proportional to $xq(x)$. Plots of that quantity are shown in Fig.~4. The lines show $x q(x)=x $  This figure shows that $x q(x,Q^2)$ does approximately oscillate about a line. This is true, in part, because including  the factor $x$ suppresses the region for which the result of using the sum of resonances, \eq{final} falls below $q(x)=1$.
We denote this   approximate oscillation  as {\it accidental duality}, because the detailed evaluations of both global and local duality expressions \eq{SR} and \eq{LSR} show a failure to achieve the necessary equality.
 
These  presented results do not  demonstrate that the use of hadronic degrees of freedom leads  to the equality of the sum, \eq{final} with the correct model quark distribution $q(x)=1$. This is expected because it is only  possible to include a finite number of excitations $X$ and because of the necessity of including a non-zero width to obtain finite values of $q(x,Q^2)$.  However,  but they do show that adding intermediate states of higher and higher masses tends toward that direction. This is because adding  states of higher mass tends to fill in gaps left by including only lower mass states. 
\section{Summary and Discussion}

Light front wave functions motivated by holographic constructions are used to study $\BGs$ in this paper.   Expressions for the structure functions  in terms of quark \eq{quark} and hadronic \eq{sum22} degrees of freedom (involving transition form factors) are presented, with an ultimate goal of obtaining a relationship between the two expressions. The specific two-parton model is defined in Sect.~IV, with masses, \eq{mass1},
 and light front wave functions,  \eq{wf},  that had been obtained in original work.  Transition form factors are expressed, using the Drell-Yan frame,  in terms of these wave functions in \eq{tf}.  
 
 The valence quark-number sum rule is presented as a new form of global duality (integral over all values of $x$ between 0 and 1) in Sect.~V, specifically in \eq{ngd} and \eq{sum23}. Using a complete set of hadronic states is necessary for this new global duality to be achieved, \eq{sum23}.  That the original work does {\it not} provide a complete set is shown in \eq{SQ}. The lack of completeness arises from the choice in the original work to examine the presumably dominant configuration of the  modes of lowest energy  in which the longitudinal wave  function of \eq{wf0} is given by $X(x)=\sqrt{x(1-x)}$. This single form is not sufficient  to provide a complete set in $x$-space. The lack of completeness is remedied by including a longitudinal potential, $U_\cL(x)$, \eq{UL}, where $\cL$ is an integer quantum number, 
  for which the stated $X(x)$ is the lowest energy mode with $\cL=0$.  The overall strength, $\l$, of $U_\cL$ is arbitrarily chosen here. In principle, this parameter, along with the longitudinal potential, should be chosen by some symmetry principle that is unknown to us. The final result should  achieve rotational invariance in the model, in the sense of having states with the proper degeneracy. This  achievement is left as a  goal of future research. Nevertheless, the present complete set is sufficient to carry out the specific purpose of this manuscript which is to examine $\BGs$.

Given the complete set of wave functions, specific expressions for the transition form factors are obtained  in \eq{ff}, \eq{FNL} and \eq{fff}.  All of the form factors fall as $1/Q^2$, \eq{limx}, at asymptotically large values. This behavior  originates in the dominance of the Feynman mechanism for this model.  That the dominant transition form factors have the same asymptotic dependence is one of the requirements to achieve $\BGs$. However, the Drell-Yan connection between form factors and structure functions is also needed, and this feature is absent in the current model.\\

The model  transition form factors are used to assess the validity of the  global, \eq{op}, and local, \eq{SR} duality sum rules. The simplicity of the  scaling quark distribution in the given model, $q(x)=1$, readily  enables studies of these sum rules, with the result that both are not satisfied within the  given model.  See Figs.~1,2.  

Evaluations of the hadronic expression for $q(x,Q^2)$, \eq{final} are presented in the previous section, see Figs.~3, 4.
The need to obtain non-infinite values to make plots of finite size  mandates that the resonant states have at least a small width, as implemented in \eq{delta}. The value of the width is chosen to be a small number $66 $ MeV. Detailed results depend upon the precise value, but the qualitative conclusions do not.
 The Figures show that including the states with $\cL>0$ are needed to approach $\BGs$, but that this duality is not obtainable with the present model. 
 
The failure to achieve $\BGs$ is not a failure of the current model because this property should be generally unexpected. Indeed, the lack of $\BGs$   teaches us an  important lesson. One might   use either hadronic or quark degrees of freedom according to  which leads to the simpler description of the specific  problem at hand. Since both sets of states are complete, it is natural to expect that   $\BGs$ should result.  The present work shows that such a  supposition is not correct. Instead, the observed validity of both global and local forms of duality for deep inelastic scattering must be related to a deeper feature of QCD.

Although the present model is very simple, it suggests a prediction that if deep inelastic scattering experiments were to be made on the pion, $\BGs$ would not be observed. This is because of the quark-anti-quark structure of the valence wave function that generally leads to a $1/Q^2$ behavior of form factors at high momentum transfer. 

Bloom \& Gilman understood their duality in terms of the asymptotic fall of resonance transition form factors, quark counting rules and a simple scaling function to represent the high $Q^2$ data.. 
The finer details of this analysis have not withstood the test of time, but measurements have shown that their duality still is viable. 
The underlying origin of this phenomenon is deeply buried within the confinement aspects of QCD. Its ultimate understanding remains a mystery.
\section*{Acknowledgements}
This work was supported by the U. S. Department of Energy Office of Science, Office of Nuclear Physics under Award Number DE-FG02-97ER- 41014 and by Battelle Memorial Institute, Pacific Northwest Division
Acting Under Contract DE-AC05-76RL01830
With the U.S. Department of Energy. Thank you to Dr. Stanley Brodsky and Dr. Guy de Teramond for the useful discussions regarding this work.  
\bibliography{duality}

\begin{thebibliography}{41}%
\makeatletter
\providecommand \@ifxundefined [1]{%
 \@ifx{#1\undefined}
}%
\providecommand \@ifnum [1]{%
 \ifnum #1\expandafter \@firstoftwo
 \else \expandafter \@secondoftwo
 \fi
}%
\providecommand \@ifx [1]{%
 \ifx #1\expandafter \@firstoftwo
 \else \expandafter \@secondoftwo
 \fi
}%
\providecommand \natexlab [1]{#1}%
\providecommand \enquote  [1]{``#1''}%
\providecommand \bibnamefont  [1]{#1}%
\providecommand \bibfnamefont [1]{#1}%
\providecommand \citenamefont [1]{#1}%
\providecommand \href@noop [0]{\@secondoftwo}%
\providecommand \href [0]{\begingroup \@sanitize@url \@href}%
\providecommand \@href[1]{\@@startlink{#1}\@@href}%
\providecommand \@@href[1]{\endgroup#1\@@endlink}%
\providecommand \@sanitize@url [0]{\catcode `\\12\catcode `\$12\catcode
  `\&12\catcode `\#12\catcode `\^12\catcode `\_12\catcode `\%12\relax}%
\providecommand \@@startlink[1]{}%
\providecommand \@@endlink[0]{}%
\providecommand \url  [0]{\begingroup\@sanitize@url \@url }%
\providecommand \@url [1]{\endgroup\@href {#1}{\urlprefix }}%
\providecommand \urlprefix  [0]{URL }%
\providecommand \Eprint [0]{\href }%
\providecommand \doibase [0]{http://dx.doi.org/}%
\providecommand \selectlanguage [0]{\@gobble}%
\providecommand \bibinfo  [0]{\@secondoftwo}%
\providecommand \bibfield  [0]{\@secondoftwo}%
\providecommand \translation [1]{[#1]}%
\providecommand \BibitemOpen [0]{}%
\providecommand \bibitemStop [0]{}%
\providecommand \bibitemNoStop [0]{.\EOS\space}%
\providecommand \EOS [0]{\spacefactor3000\relax}%
\providecommand \BibitemShut  [1]{\csname bibitem#1\endcsname}%
\let\auto@bib@innerbib\@empty
\bibitem [{\citenamefont {Melnitchouk}\ \emph {et~al.}(2005)\citenamefont
  {Melnitchouk}, \citenamefont {Ent},\ and\ \citenamefont
  {Keppel}}]{Melnitchouk:2005zr}%
  \BibitemOpen
  \bibfield  {author} {\bibinfo {author} {\bibfnamefont {W.}~\bibnamefont
  {Melnitchouk}}, \bibinfo {author} {\bibfnamefont {R.}~\bibnamefont {Ent}}, \
  and\ \bibinfo {author} {\bibfnamefont {C.}~\bibnamefont {Keppel}},\
  }\bibfield  {title} {\enquote {\bibinfo {title} {{Quark-hadron duality in
  electron scattering}},}\ }\href {\doibase 10.1016/j.physrep.2004.10.004}
  {\bibfield  {journal} {\bibinfo  {journal} {Phys. Rept.}\ }\textbf {\bibinfo
  {volume} {406}},\ \bibinfo {pages} {127--301} (\bibinfo {year} {2005})},\
  \Eprint {http://arxiv.org/abs/hep-ph/0501217} {arXiv:hep-ph/0501217}
  \BibitemShut {NoStop}%
\bibitem [{\citenamefont {Bloom}\ and\ \citenamefont
  {Gilman}(1970)}]{Bloom:1970xb}%
  \BibitemOpen
  \bibfield  {author} {\bibinfo {author} {\bibfnamefont {Elliott~D.}\
  \bibnamefont {Bloom}}\ and\ \bibinfo {author} {\bibfnamefont {Frederick~J.}\
  \bibnamefont {Gilman}},\ }\bibfield  {title} {\enquote {\bibinfo {title}
  {{Scaling, Duality, and the Behavior of Resonances in Inelastic
  electron-Proton Scattering}},}\ }\href {\doibase 10.1103/PhysRevLett.25.1140}
  {\bibfield  {journal} {\bibinfo  {journal} {Phys. Rev. Lett.}\ }\textbf
  {\bibinfo {volume} {25}},\ \bibinfo {pages} {1140} (\bibinfo {year}
  {1970})}\BibitemShut {NoStop}%
\bibitem [{\citenamefont {Bloom}\ and\ \citenamefont
  {Gilman}(1971)}]{Bloom:1971ye}%
  \BibitemOpen
  \bibfield  {author} {\bibinfo {author} {\bibfnamefont {Elliott~D.}\
  \bibnamefont {Bloom}}\ and\ \bibinfo {author} {\bibfnamefont {Frederick~J.}\
  \bibnamefont {Gilman}},\ }\bibfield  {title} {\enquote {\bibinfo {title}
  {{Scaling and the Behavior of Nucleon Resonances in Inelastic
  electron-Nucleon Scattering}},}\ }\href {\doibase 10.1103/PhysRevD.4.2901}
  {\bibfield  {journal} {\bibinfo  {journal} {Phys. Rev. D}\ }\textbf {\bibinfo
  {volume} {4}},\ \bibinfo {pages} {2901} (\bibinfo {year} {1971})}\BibitemShut
  {NoStop}%
\bibitem [{\citenamefont {De~Rujula}\ \emph {et~al.}(1976)\citenamefont
  {De~Rujula}, \citenamefont {Georgi},\ and\ \citenamefont
  {Politzer}}]{DeRujula:1976ke}%
  \BibitemOpen
  \bibfield  {author} {\bibinfo {author} {\bibfnamefont {A.}~\bibnamefont
  {De~Rujula}}, \bibinfo {author} {\bibfnamefont {Howard}\ \bibnamefont
  {Georgi}}, \ and\ \bibinfo {author} {\bibfnamefont {H.David}\ \bibnamefont
  {Politzer}},\ }\bibfield  {title} {\enquote {\bibinfo {title} {{An
  Explanation of Local Duality and Precocious Scaling}},}\ }\href {\doibase
  10.1016/0370-2693(76)90113-1} {\bibfield  {journal} {\bibinfo  {journal}
  {Phys. Lett. B}\ }\textbf {\bibinfo {volume} {64}},\ \bibinfo {pages}
  {428--432} (\bibinfo {year} {1976})}\BibitemShut {NoStop}%
\bibitem [{\citenamefont {De~Rujula}\ \emph {et~al.}(1977)\citenamefont
  {De~Rujula}, \citenamefont {Georgi},\ and\ \citenamefont
  {Politzer}}]{DeRujula:1976baf}%
  \BibitemOpen
  \bibfield  {author} {\bibinfo {author} {\bibfnamefont {Alvaro}\ \bibnamefont
  {De~Rujula}}, \bibinfo {author} {\bibfnamefont {Howard}\ \bibnamefont
  {Georgi}}, \ and\ \bibinfo {author} {\bibfnamefont {H.David}\ \bibnamefont
  {Politzer}},\ }\bibfield  {title} {\enquote {\bibinfo {title}
  {{Demythification of Electroproduction, Local Duality and Precocious
  Scaling}},}\ }\href {\doibase 10.1016/S0003-4916(97)90003-8} {\bibfield
  {journal} {\bibinfo  {journal} {Annals Phys.}\ }\textbf {\bibinfo {volume}
  {103}},\ \bibinfo {pages} {315} (\bibinfo {year} {1977})}\BibitemShut
  {NoStop}%
\bibitem [{\citenamefont {Niculescu}\ \emph
  {et~al.}(2000{\natexlab{a}})\citenamefont {Niculescu} \emph
  {et~al.}}]{Niculescu:2000tk}%
  \BibitemOpen
  \bibfield  {author} {\bibinfo {author} {\bibfnamefont {I.}~\bibnamefont
  {Niculescu}} \emph {et~al.},\ }\bibfield  {title} {\enquote {\bibinfo {title}
  {{Experimental verification of quark hadron duality}},}\ }\href {\doibase
  10.1103/PhysRevLett.85.1186} {\bibfield  {journal} {\bibinfo  {journal}
  {Phys. Rev. Lett.}\ }\textbf {\bibinfo {volume} {85}},\ \bibinfo {pages}
  {1186--1189} (\bibinfo {year} {2000}{\natexlab{a}})}\BibitemShut {NoStop}%
\bibitem [{\citenamefont {Niculescu}\ \emph
  {et~al.}(2000{\natexlab{b}})\citenamefont {Niculescu} \emph
  {et~al.}}]{Niculescu:2000tj}%
  \BibitemOpen
  \bibfield  {author} {\bibinfo {author} {\bibfnamefont {I.}~\bibnamefont
  {Niculescu}} \emph {et~al.},\ }\bibfield  {title} {\enquote {\bibinfo {title}
  {{Evidence for valencelike quark hadron duality}},}\ }\href {\doibase
  10.1103/PhysRevLett.85.1182} {\bibfield  {journal} {\bibinfo  {journal}
  {Phys. Rev. Lett.}\ }\textbf {\bibinfo {volume} {85}},\ \bibinfo {pages}
  {1182--1185} (\bibinfo {year} {2000}{\natexlab{b}})}\BibitemShut {NoStop}%
\bibitem [{\citenamefont {Ent}\ \emph {et~al.}(2000)\citenamefont {Ent},
  \citenamefont {Keppel},\ and\ \citenamefont {Niculescu}}]{Ent:2000jj}%
  \BibitemOpen
  \bibfield  {author} {\bibinfo {author} {\bibfnamefont {R.}~\bibnamefont
  {Ent}}, \bibinfo {author} {\bibfnamefont {C.E.}\ \bibnamefont {Keppel}}, \
  and\ \bibinfo {author} {\bibfnamefont {I.}~\bibnamefont {Niculescu}},\
  }\bibfield  {title} {\enquote {\bibinfo {title} {{Nucleon elastic
  form-factors and local duality}},}\ }\href {\doibase
  10.1103/PhysRevD.62.073008} {\bibfield  {journal} {\bibinfo  {journal} {Phys.
  Rev. D}\ }\textbf {\bibinfo {volume} {62}},\ \bibinfo {pages} {073008}
  (\bibinfo {year} {2000})}\BibitemShut {NoStop}%
\bibitem [{\citenamefont {Einhorn}(1976)}]{PhysRevD.14.3451}%
  \BibitemOpen
  \bibfield  {author} {\bibinfo {author} {\bibfnamefont {Martin~B.}\
  \bibnamefont {Einhorn}},\ }\bibfield  {title} {\enquote {\bibinfo {title}
  {Confinement, form factors, and deep-inelastic scattering in two-dimensional
  quantum chromodynamics},}\ }\href {\doibase 10.1103/PhysRevD.14.3451}
  {\bibfield  {journal} {\bibinfo  {journal} {Phys. Rev. D}\ }\textbf {\bibinfo
  {volume} {14}},\ \bibinfo {pages} {3451--3471} (\bibinfo {year}
  {1976})}\BibitemShut {NoStop}%
\bibitem [{\citenamefont {Domokos}\ \emph
  {et~al.}(1971{\natexlab{a}})\citenamefont {Domokos}, \citenamefont
  {Kovesi-Domokos},\ and\ \citenamefont {Schonberg}}]{Domokos:1971ds}%
  \BibitemOpen
  \bibfield  {author} {\bibinfo {author} {\bibfnamefont {G.}~\bibnamefont
  {Domokos}}, \bibinfo {author} {\bibfnamefont {S.}~\bibnamefont
  {Kovesi-Domokos}}, \ and\ \bibinfo {author} {\bibfnamefont {E.}~\bibnamefont
  {Schonberg}},\ }\bibfield  {title} {\enquote {\bibinfo {title}
  {{Direct-channel resonance model of deep-inelastic electron scattering. i.
  scattering on unpolarized targets}},}\ }\href {\doibase
  10.1103/PhysRevD.3.1184} {\bibfield  {journal} {\bibinfo  {journal} {Phys.
  Rev. D}\ }\textbf {\bibinfo {volume} {3}},\ \bibinfo {pages} {1184--1190}
  (\bibinfo {year} {1971}{\natexlab{a}})}\BibitemShut {NoStop}%
\bibitem [{\citenamefont {Domokos}\ \emph
  {et~al.}(1971{\natexlab{b}})\citenamefont {Domokos}, \citenamefont
  {Kovesi-Domokos},\ and\ \citenamefont {Schonberg}}]{Domokos:1972yc}%
  \BibitemOpen
  \bibfield  {author} {\bibinfo {author} {\bibfnamefont {G.}~\bibnamefont
  {Domokos}}, \bibinfo {author} {\bibfnamefont {S.}~\bibnamefont
  {Kovesi-Domokos}}, \ and\ \bibinfo {author} {\bibfnamefont {E.}~\bibnamefont
  {Schonberg}},\ }\bibfield  {title} {\enquote {\bibinfo {title}
  {{Deep-inelastic neutrino scattering in a resonance model}},}\ }\href
  {\doibase 10.1103/PhysRevD.4.2115} {\bibfield  {journal} {\bibinfo  {journal}
  {Phys. Rev. D}\ }\textbf {\bibinfo {volume} {4}},\ \bibinfo {pages}
  {2115--2125} (\bibinfo {year} {1971}{\natexlab{b}})}\BibitemShut {NoStop}%
\bibitem [{\citenamefont {Isgur}\ \emph {et~al.}(2001)\citenamefont {Isgur},
  \citenamefont {Jeschonnek}, \citenamefont {Melnitchouk},\ and\ \citenamefont
  {Van~Orden}}]{Isgur:2001bt}%
  \BibitemOpen
  \bibfield  {author} {\bibinfo {author} {\bibfnamefont {Nathan}\ \bibnamefont
  {Isgur}}, \bibinfo {author} {\bibfnamefont {Sabine.}\ \bibnamefont
  {Jeschonnek}}, \bibinfo {author} {\bibfnamefont {W.}~\bibnamefont
  {Melnitchouk}}, \ and\ \bibinfo {author} {\bibfnamefont {J.W.}\ \bibnamefont
  {Van~Orden}},\ }\bibfield  {title} {\enquote {\bibinfo {title} {{Quark hadron
  duality in structure functions}},}\ }\href {\doibase
  10.1103/PhysRevD.64.054005} {\bibfield  {journal} {\bibinfo  {journal} {Phys.
  Rev. D}\ }\textbf {\bibinfo {volume} {64}},\ \bibinfo {pages} {054005}
  (\bibinfo {year} {2001})},\ \Eprint {http://arxiv.org/abs/hep-ph/0104022}
  {arXiv:hep-ph/0104022} \BibitemShut {NoStop}%
\bibitem [{\citenamefont {Jeschonnek}\ and\ \citenamefont
  {Van~Orden}(2004)}]{Jeschonnek:2003sb}%
  \BibitemOpen
  \bibfield  {author} {\bibinfo {author} {\bibfnamefont {Sabine}\ \bibnamefont
  {Jeschonnek}}\ and\ \bibinfo {author} {\bibfnamefont {J.W.}\ \bibnamefont
  {Van~Orden}},\ }\bibfield  {title} {\enquote {\bibinfo {title} {{Modeling
  quark hadron duality for relativistic, confined fermions}},}\ }\href
  {\doibase 10.1103/PhysRevD.69.054006} {\bibfield  {journal} {\bibinfo
  {journal} {Phys. Rev. D}\ }\textbf {\bibinfo {volume} {69}},\ \bibinfo
  {pages} {054006} (\bibinfo {year} {2004})},\ \Eprint
  {http://arxiv.org/abs/hep-ph/0310298} {arXiv:hep-ph/0310298} \BibitemShut
  {NoStop}%
\bibitem [{\citenamefont {Close}\ and\ \citenamefont
  {Zhao}(2002)}]{Close:2002tm}%
  \BibitemOpen
  \bibfield  {author} {\bibinfo {author} {\bibfnamefont {Frank~E.}\
  \bibnamefont {Close}}\ and\ \bibinfo {author} {\bibfnamefont {Qiang}\
  \bibnamefont {Zhao}},\ }\bibfield  {title} {\enquote {\bibinfo {title} {{A
  Pedagogic model for deeply virtual Compton scattering with quark hadron
  duality}},}\ }\href {\doibase 10.1103/PhysRevD.66.054001} {\bibfield
  {journal} {\bibinfo  {journal} {Phys. Rev. D}\ }\textbf {\bibinfo {volume}
  {66}},\ \bibinfo {pages} {054001} (\bibinfo {year} {2002})},\ \Eprint
  {http://arxiv.org/abs/hep-ph/0202181} {arXiv:hep-ph/0202181} \BibitemShut
  {NoStop}%
\bibitem [{\citenamefont {Harrington}(2002)}]{PhysRevC.66.065205}%
  \BibitemOpen
  \bibfield  {author} {\bibinfo {author} {\bibfnamefont {David~R.}\
  \bibnamefont {Harrington}},\ }\bibfield  {title} {\enquote {\bibinfo {title}
  {Asymptotic freedom for nonrelativistic confinement},}\ }\href {\doibase
  10.1103/PhysRevC.66.065205} {\bibfield  {journal} {\bibinfo  {journal} {Phys.
  Rev. C}\ }\textbf {\bibinfo {volume} {66}},\ \bibinfo {pages} {065205}
  (\bibinfo {year} {2002})}\BibitemShut {NoStop}%
\bibitem [{\citenamefont {Close}\ and\ \citenamefont
  {Isgur}(2001)}]{Close:2001ha}%
  \BibitemOpen
  \bibfield  {author} {\bibinfo {author} {\bibfnamefont {Frank~E.}\
  \bibnamefont {Close}}\ and\ \bibinfo {author} {\bibfnamefont {Nathan}\
  \bibnamefont {Isgur}},\ }\bibfield  {title} {\enquote {\bibinfo {title} {{The
  Origins of quark hadron duality: How does the square of the sum become the
  sum of the squares?}}}\ }\href {\doibase 10.1016/S0370-2693(01)00548-2}
  {\bibfield  {journal} {\bibinfo  {journal} {Phys. Lett. B}\ }\textbf
  {\bibinfo {volume} {509}},\ \bibinfo {pages} {81--86} (\bibinfo {year}
  {2001})},\ \Eprint {http://arxiv.org/abs/hep-ph/0102067}
  {arXiv:hep-ph/0102067} \BibitemShut {NoStop}%
\bibitem [{\citenamefont {Paris}\ and\ \citenamefont
  {Pandharipande}(2001)}]{Paris:2001pm}%
  \BibitemOpen
  \bibfield  {author} {\bibinfo {author} {\bibfnamefont {M.W.}\ \bibnamefont
  {Paris}}\ and\ \bibinfo {author} {\bibfnamefont {Vijay~R.}\ \bibnamefont
  {Pandharipande}},\ }\bibfield  {title} {\enquote {\bibinfo {title} {{Scaling
  of space and time - like response of confined relativistic particles}},}\
  }\href {\doibase 10.1016/S0370-2693(01)00794-8} {\bibfield  {journal}
  {\bibinfo  {journal} {Phys. Lett. B}\ }\textbf {\bibinfo {volume} {514}},\
  \bibinfo {pages} {361--365} (\bibinfo {year} {2001})},\ \Eprint
  {http://arxiv.org/abs/nucl-th/0105076} {arXiv:nucl-th/0105076} \BibitemShut
  {NoStop}%
\bibitem [{\citenamefont {Paris}\ and\ \citenamefont
  {Pandharipande}(2002)}]{PhysRevC.65.035203}%
  \BibitemOpen
  \bibfield  {author} {\bibinfo {author} {\bibfnamefont {Mark~W.}\ \bibnamefont
  {Paris}}\ and\ \bibinfo {author} {\bibfnamefont {Vijay~R.}\ \bibnamefont
  {Pandharipande}},\ }\bibfield  {title} {\enquote {\bibinfo {title} {Final
  state interaction contribution to the response of confined relativistic
  particles},}\ }\href {\doibase 10.1103/PhysRevC.65.035203} {\bibfield
  {journal} {\bibinfo  {journal} {Phys. Rev. C}\ }\textbf {\bibinfo {volume}
  {65}},\ \bibinfo {pages} {035203} (\bibinfo {year} {2002})}\BibitemShut
  {NoStop}%
\bibitem [{\citenamefont {Brodsky}\ \emph {et~al.}(2015)\citenamefont
  {Brodsky}, \citenamefont {de~Teramond}, \citenamefont {Dosch},\ and\
  \citenamefont {Erlich}}]{Brodsky:2014yha}%
  \BibitemOpen
  \bibfield  {author} {\bibinfo {author} {\bibfnamefont {Stanley~J.}\
  \bibnamefont {Brodsky}}, \bibinfo {author} {\bibfnamefont {Guy~F.}\
  \bibnamefont {de~Teramond}}, \bibinfo {author} {\bibfnamefont {Hans~Gunter}\
  \bibnamefont {Dosch}}, \ and\ \bibinfo {author} {\bibfnamefont {Joshua}\
  \bibnamefont {Erlich}},\ }\bibfield  {title} {\enquote {\bibinfo {title}
  {{Light-Front Holographic QCD and Emerging Confinement}},}\ }\href {\doibase
  10.1016/j.physrep.2015.05.001} {\bibfield  {journal} {\bibinfo  {journal}
  {Phys. Rept.}\ }\textbf {\bibinfo {volume} {584}},\ \bibinfo {pages} {1--105}
  (\bibinfo {year} {2015})},\ \Eprint {http://arxiv.org/abs/1407.8131}
  {arXiv:1407.8131 [hep-ph]} \BibitemShut {NoStop}%
\bibitem [{\citenamefont {Brodsky}\ \emph {et~al.}(1998)\citenamefont
  {Brodsky}, \citenamefont {Pauli},\ and\ \citenamefont
  {Pinsky}}]{Brodsky:1997de}%
  \BibitemOpen
  \bibfield  {author} {\bibinfo {author} {\bibfnamefont {Stanley~J.}\
  \bibnamefont {Brodsky}}, \bibinfo {author} {\bibfnamefont {Hans-Christian}\
  \bibnamefont {Pauli}}, \ and\ \bibinfo {author} {\bibfnamefont {Stephen~S.}\
  \bibnamefont {Pinsky}},\ }\bibfield  {title} {\enquote {\bibinfo {title}
  {{Quantum chromodynamics and other field theories on the light cone}},}\
  }\href {\doibase 10.1016/S0370-1573(97)00089-6} {\bibfield  {journal}
  {\bibinfo  {journal} {Phys. Rept.}\ }\textbf {\bibinfo {volume} {301}},\
  \bibinfo {pages} {299--486} (\bibinfo {year} {1998})},\ \Eprint
  {http://arxiv.org/abs/hep-ph/9705477} {arXiv:hep-ph/9705477} \BibitemShut
  {NoStop}%
\bibitem [{\citenamefont {Dirac}(1949)}]{Dirac:1949cp}%
  \BibitemOpen
  \bibfield  {author} {\bibinfo {author} {\bibfnamefont {Paul~A.M.}\
  \bibnamefont {Dirac}},\ }\bibfield  {title} {\enquote {\bibinfo {title}
  {{Forms of Relativistic Dynamics}},}\ }\href {\doibase
  10.1103/RevModPhys.21.392} {\bibfield  {journal} {\bibinfo  {journal} {Rev.
  Mod. Phys.}\ }\textbf {\bibinfo {volume} {21}},\ \bibinfo {pages} {392--399}
  (\bibinfo {year} {1949})}\BibitemShut {NoStop}%
\bibitem [{\citenamefont {de~Teramond}\ and\ \citenamefont
  {Brodsky}(2009)}]{deTeramond:2008ht}%
  \BibitemOpen
  \bibfield  {author} {\bibinfo {author} {\bibfnamefont {Guy~F.}\ \bibnamefont
  {de~Teramond}}\ and\ \bibinfo {author} {\bibfnamefont {Stanley~J.}\
  \bibnamefont {Brodsky}},\ }\bibfield  {title} {\enquote {\bibinfo {title}
  {{Light-Front Holography: A First Approximation to QCD}},}\ }\href {\doibase
  10.1103/PhysRevLett.102.081601} {\bibfield  {journal} {\bibinfo  {journal}
  {Phys. Rev. Lett.}\ }\textbf {\bibinfo {volume} {102}},\ \bibinfo {pages}
  {081601} (\bibinfo {year} {2009})},\ \Eprint {http://arxiv.org/abs/0809.4899}
  {arXiv:0809.4899 [hep-ph]} \BibitemShut {NoStop}%
\bibitem [{\citenamefont {Brodsky}\ and\ \citenamefont
  {de~Teramond}(2006)}]{Brodsky:2006uqa}%
  \BibitemOpen
  \bibfield  {author} {\bibinfo {author} {\bibfnamefont {Stanley~J.}\
  \bibnamefont {Brodsky}}\ and\ \bibinfo {author} {\bibfnamefont {Guy~F.}\
  \bibnamefont {de~Teramond}},\ }\bibfield  {title} {\enquote {\bibinfo {title}
  {{Hadronic spectra and light-front wavefunctions in holographic QCD}},}\
  }\href {\doibase 10.1103/PhysRevLett.96.201601} {\bibfield  {journal}
  {\bibinfo  {journal} {Phys. Rev. Lett.}\ }\textbf {\bibinfo {volume} {96}},\
  \bibinfo {pages} {201601} (\bibinfo {year} {2006})},\ \Eprint
  {http://arxiv.org/abs/hep-ph/0602252} {arXiv:hep-ph/0602252} \BibitemShut
  {NoStop}%
\bibitem [{\citenamefont {Brodsky}\ and\ \citenamefont
  {de~Teramond}(2008{\natexlab{a}})}]{Brodsky:2007hb}%
  \BibitemOpen
  \bibfield  {author} {\bibinfo {author} {\bibfnamefont {Stanley~J.}\
  \bibnamefont {Brodsky}}\ and\ \bibinfo {author} {\bibfnamefont {Guy~F.}\
  \bibnamefont {de~Teramond}},\ }\bibfield  {title} {\enquote {\bibinfo {title}
  {{Light-Front Dynamics and AdS/QCD Correspondence: The Pion Form Factor in
  the Space- and Time-Like Regions}},}\ }\href {\doibase
  10.1103/PhysRevD.77.056007} {\bibfield  {journal} {\bibinfo  {journal} {Phys.
  Rev. D}\ }\textbf {\bibinfo {volume} {77}},\ \bibinfo {pages} {056007}
  (\bibinfo {year} {2008}{\natexlab{a}})},\ \Eprint
  {http://arxiv.org/abs/0707.3859} {arXiv:0707.3859 [hep-ph]} \BibitemShut
  {NoStop}%
\bibitem [{\citenamefont {Polchinski}\ and\ \citenamefont
  {Strassler}(2003)}]{Polchinski:2002jw}%
  \BibitemOpen
  \bibfield  {author} {\bibinfo {author} {\bibfnamefont {Joseph}\ \bibnamefont
  {Polchinski}}\ and\ \bibinfo {author} {\bibfnamefont {Matthew~J.}\
  \bibnamefont {Strassler}},\ }\bibfield  {title} {\enquote {\bibinfo {title}
  {{Deep inelastic scattering and gauge / string duality}},}\ }\href {\doibase
  10.1088/1126-6708/2003/05/012} {\bibfield  {journal} {\bibinfo  {journal}
  {JHEP}\ }\textbf {\bibinfo {volume} {05}},\ \bibinfo {pages} {012} (\bibinfo
  {year} {2003})},\ \Eprint {http://arxiv.org/abs/hep-th/0209211}
  {arXiv:hep-th/0209211} \BibitemShut {NoStop}%
\bibitem [{\citenamefont {Drell}\ and\ \citenamefont
  {Yan}(1970{\natexlab{a}})}]{Drell:1969km}%
  \BibitemOpen
  \bibfield  {author} {\bibinfo {author} {\bibfnamefont {S.D.}\ \bibnamefont
  {Drell}}\ and\ \bibinfo {author} {\bibfnamefont {Tung-Mow}\ \bibnamefont
  {Yan}},\ }\bibfield  {title} {\enquote {\bibinfo {title} {{Connection of
  Elastic Electromagnetic Nucleon Form-Factors at Large Q**2 and Deep Inelastic
  Structure Functions Near Threshold}},}\ }\href {\doibase
  10.1103/PhysRevLett.24.181} {\bibfield  {journal} {\bibinfo  {journal} {Phys.
  Rev. Lett.}\ }\textbf {\bibinfo {volume} {24}},\ \bibinfo {pages} {181--185}
  (\bibinfo {year} {1970}{\natexlab{a}})}\BibitemShut {NoStop}%
\bibitem [{\citenamefont {West}(1970)}]{West:1970av}%
  \BibitemOpen
  \bibfield  {author} {\bibinfo {author} {\bibfnamefont {Geoffrey~B.}\
  \bibnamefont {West}},\ }\bibfield  {title} {\enquote {\bibinfo {title}
  {{Phenomenological model for the electromagnetic structure of the proton}},}\
  }\href {\doibase 10.1103/PhysRevLett.24.1206} {\bibfield  {journal} {\bibinfo
   {journal} {Phys. Rev. Lett.}\ }\textbf {\bibinfo {volume} {24}},\ \bibinfo
  {pages} {1206--1209} (\bibinfo {year} {1970})}\BibitemShut {NoStop}%
\bibitem [{\citenamefont {Brodsky}\ and\ \citenamefont
  {de~Teramond}(2008{\natexlab{b}})}]{Brodsky:2008pf}%
  \BibitemOpen
  \bibfield  {author} {\bibinfo {author} {\bibfnamefont {Stanley~J.}\
  \bibnamefont {Brodsky}}\ and\ \bibinfo {author} {\bibfnamefont {Guy~F.}\
  \bibnamefont {de~Teramond}},\ }\bibfield  {title} {\enquote {\bibinfo {title}
  {{Light-Front Dynamics and AdS/QCD Correspondence: Gravitational Form Factors
  of Composite Hadrons}},}\ }\href {\doibase 10.1103/PhysRevD.78.025032}
  {\bibfield  {journal} {\bibinfo  {journal} {Phys. Rev. D}\ }\textbf {\bibinfo
  {volume} {78}},\ \bibinfo {pages} {025032} (\bibinfo {year}
  {2008}{\natexlab{b}})},\ \Eprint {http://arxiv.org/abs/0804.0452}
  {arXiv:0804.0452 [hep-ph]} \BibitemShut {NoStop}%
\bibitem [{\citenamefont {Abidin}\ and\ \citenamefont
  {Carlson}(2008)}]{Abidin:2008ku}%
  \BibitemOpen
  \bibfield  {author} {\bibinfo {author} {\bibfnamefont {Zainul}\ \bibnamefont
  {Abidin}}\ and\ \bibinfo {author} {\bibfnamefont {Carl~E.}\ \bibnamefont
  {Carlson}},\ }\bibfield  {title} {\enquote {\bibinfo {title} {{Gravitational
  form factors of vector mesons in an AdS/QCD model}},}\ }\href {\doibase
  10.1103/PhysRevD.77.095007} {\bibfield  {journal} {\bibinfo  {journal} {Phys.
  Rev. D}\ }\textbf {\bibinfo {volume} {77}},\ \bibinfo {pages} {095007}
  (\bibinfo {year} {2008})},\ \Eprint {http://arxiv.org/abs/0801.3839}
  {arXiv:0801.3839 [hep-ph]} \BibitemShut {NoStop}%
\bibitem [{\citenamefont {Drell}\ and\ \citenamefont
  {Yan}(1970{\natexlab{b}})}]{PhysRevLett.24.181}%
  \BibitemOpen
  \bibfield  {author} {\bibinfo {author} {\bibfnamefont {Sidney~D.}\
  \bibnamefont {Drell}}\ and\ \bibinfo {author} {\bibfnamefont {Tung-Mow}\
  \bibnamefont {Yan}},\ }\bibfield  {title} {\enquote {\bibinfo {title}
  {Connection of elastic electromagnetic nucleon form factors at large
  ${Q}^{2}$ and deep inelastic structure functions near threshold},}\ }\href
  {\doibase 10.1103/PhysRevLett.24.181} {\bibfield  {journal} {\bibinfo
  {journal} {Phys. Rev. Lett.}\ }\textbf {\bibinfo {volume} {24}},\ \bibinfo
  {pages} {181--186} (\bibinfo {year} {1970}{\natexlab{b}})}\BibitemShut
  {NoStop}%
\bibitem [{\citenamefont {Karch}\ \emph {et~al.}(2006)\citenamefont {Karch},
  \citenamefont {Katz}, \citenamefont {Son},\ and\ \citenamefont
  {Stephanov}}]{Karch:2006pv}%
  \BibitemOpen
  \bibfield  {author} {\bibinfo {author} {\bibfnamefont {Andreas}\ \bibnamefont
  {Karch}}, \bibinfo {author} {\bibfnamefont {Emanuel}\ \bibnamefont {Katz}},
  \bibinfo {author} {\bibfnamefont {Dam~T.}\ \bibnamefont {Son}}, \ and\
  \bibinfo {author} {\bibfnamefont {Mikhail~A.}\ \bibnamefont {Stephanov}},\
  }\bibfield  {title} {\enquote {\bibinfo {title} {{Linear confinement and
  AdS/QCD}},}\ }\href {\doibase 10.1103/PhysRevD.74.015005} {\bibfield
  {journal} {\bibinfo  {journal} {Phys. Rev. D}\ }\textbf {\bibinfo {volume}
  {74}},\ \bibinfo {pages} {015005} (\bibinfo {year} {2006})},\ \Eprint
  {http://arxiv.org/abs/hep-ph/0602229} {arXiv:hep-ph/0602229} \BibitemShut
  {NoStop}%
\bibitem [{\citenamefont {de~T\'eramond}\ and\ \citenamefont
  {Brodsky}(2009)}]{PhysRevLett.102.081601}%
  \BibitemOpen
  \bibfield  {author} {\bibinfo {author} {\bibfnamefont {Guy~F.}\ \bibnamefont
  {de~T\'eramond}}\ and\ \bibinfo {author} {\bibfnamefont {Stanley~J.}\
  \bibnamefont {Brodsky}},\ }\bibfield  {title} {\enquote {\bibinfo {title}
  {Light-front holography: A first approximation to qcd},}\ }\href {\doibase
  10.1103/PhysRevLett.102.081601} {\bibfield  {journal} {\bibinfo  {journal}
  {Phys. Rev. Lett.}\ }\textbf {\bibinfo {volume} {102}},\ \bibinfo {pages}
  {081601} (\bibinfo {year} {2009})}\BibitemShut {NoStop}%
\bibitem [{\citenamefont {de~T\'eramond}\ \emph {et~al.}(2015)\citenamefont
  {de~T\'eramond}, \citenamefont {Dosch},\ and\ \citenamefont
  {Brodsky}}]{PhysRevD.91.045040}%
  \BibitemOpen
  \bibfield  {author} {\bibinfo {author} {\bibfnamefont {Guy~F.}\ \bibnamefont
  {de~T\'eramond}}, \bibinfo {author} {\bibfnamefont {Hans~G\"unter}\
  \bibnamefont {Dosch}}, \ and\ \bibinfo {author} {\bibfnamefont {Stanley~J.}\
  \bibnamefont {Brodsky}},\ }\bibfield  {title} {\enquote {\bibinfo {title}
  {Baryon spectrum from superconformal quantum mechanics and its light-front
  holographic embedding},}\ }\href {\doibase 10.1103/PhysRevD.91.045040}
  {\bibfield  {journal} {\bibinfo  {journal} {Phys. Rev. D}\ }\textbf {\bibinfo
  {volume} {91}},\ \bibinfo {pages} {045040} (\bibinfo {year}
  {2015})}\BibitemShut {NoStop}%
\bibitem [{\citenamefont {Dosch}\ \emph {et~al.}(2015)\citenamefont {Dosch},
  \citenamefont {de~T\'eramond},\ and\ \citenamefont
  {Brodsky}}]{PhysRevD.91.085016}%
  \BibitemOpen
  \bibfield  {author} {\bibinfo {author} {\bibfnamefont {Hans~G\"unter}\
  \bibnamefont {Dosch}}, \bibinfo {author} {\bibfnamefont {Guy~F.}\
  \bibnamefont {de~T\'eramond}}, \ and\ \bibinfo {author} {\bibfnamefont
  {Stanley~J.}\ \bibnamefont {Brodsky}},\ }\bibfield  {title} {\enquote
  {\bibinfo {title} {Superconformal baryon-meson symmetry and light-front
  holographic qcd},}\ }\href {\doibase 10.1103/PhysRevD.91.085016} {\bibfield
  {journal} {\bibinfo  {journal} {Phys. Rev. D}\ }\textbf {\bibinfo {volume}
  {91}},\ \bibinfo {pages} {085016} (\bibinfo {year} {2015})}\BibitemShut
  {NoStop}%
\bibitem [{\citenamefont {Lepage}\ and\ \citenamefont
  {Brodsky}(1980)}]{Lepage:1980fj}%
  \BibitemOpen
  \bibfield  {author} {\bibinfo {author} {\bibfnamefont {G.Peter}\ \bibnamefont
  {Lepage}}\ and\ \bibinfo {author} {\bibfnamefont {Stanley~J.}\ \bibnamefont
  {Brodsky}},\ }\bibfield  {title} {\enquote {\bibinfo {title} {{Exclusive
  Processes in Perturbative Quantum Chromodynamics}},}\ }\href {\doibase
  10.1103/PhysRevD.22.2157} {\bibfield  {journal} {\bibinfo  {journal} {Phys.
  Rev. D}\ }\textbf {\bibinfo {volume} {22}},\ \bibinfo {pages} {2157}
  (\bibinfo {year} {1980})}\BibitemShut {NoStop}%
\bibitem [{\citenamefont {Li}\ \emph {et~al.}(2016)\citenamefont {Li},
  \citenamefont {Maris}, \citenamefont {Zhao},\ and\ \citenamefont
  {Vary}}]{Li:2015zda}%
  \BibitemOpen
  \bibfield  {author} {\bibinfo {author} {\bibfnamefont {Yang}\ \bibnamefont
  {Li}}, \bibinfo {author} {\bibfnamefont {Pieter}\ \bibnamefont {Maris}},
  \bibinfo {author} {\bibfnamefont {Xingbo}\ \bibnamefont {Zhao}}, \ and\
  \bibinfo {author} {\bibfnamefont {James~P.}\ \bibnamefont {Vary}},\
  }\bibfield  {title} {\enquote {\bibinfo {title} {{Heavy Quarkonium in a
  Holographic Basis}},}\ }\href {\doibase 10.1016/j.physletb.2016.04.065}
  {\bibfield  {journal} {\bibinfo  {journal} {Phys. Lett. B}\ }\textbf
  {\bibinfo {volume} {758}},\ \bibinfo {pages} {118--124} (\bibinfo {year}
  {2016})},\ \Eprint {http://arxiv.org/abs/1509.07212} {arXiv:1509.07212
  [hep-ph]} \BibitemShut {NoStop}%
\bibitem [{\citenamefont {Miller}\ and\ \citenamefont
  {Brodsky}(2020)}]{Miller:2019ysh}%
  \BibitemOpen
  \bibfield  {author} {\bibinfo {author} {\bibfnamefont {Gerald~A.}\
  \bibnamefont {Miller}}\ and\ \bibinfo {author} {\bibfnamefont {Stanley~J.}\
  \bibnamefont {Brodsky}},\ }\bibfield  {title} {\enquote {\bibinfo {title}
  {{Frame-independent spatial coordinate $\tilde{z}$: Implications for
  light-front wave functions, deep inelastic scattering, light-front
  holography, and lattice QCD calculations}},}\ }\href {\doibase
  10.1103/PhysRevC.102.022201} {\bibfield  {journal} {\bibinfo  {journal}
  {Phys. Rev. C}\ }\textbf {\bibinfo {volume} {102}},\ \bibinfo {pages}
  {022201} (\bibinfo {year} {2020})},\ \Eprint
  {http://arxiv.org/abs/1912.08911} {arXiv:1912.08911 [hep-ph]} \BibitemShut
  {NoStop}%
\bibitem [{\citenamefont {Feynman}(1972)}]{Feynman:1973xc}%
  \BibitemOpen
  \bibfield  {author} {\bibinfo {author} {\bibfnamefont {R.P.}\ \bibnamefont
  {Feynman}},\ }\href@noop {} {\emph {\bibinfo {title} {{Photon-hadron
  interactions}}}}\ (\bibinfo  {publisher} {W.A. Benjamin, Inc.},\ \bibinfo
  {address} {Reading},\ \bibinfo {year} {1972})\BibitemShut {NoStop}%
\bibitem [{\citenamefont {Brodsky}\ \emph {et~al.}(1995)\citenamefont
  {Brodsky}, \citenamefont {Burkardt},\ and\ \citenamefont
  {Schmidt}}]{Brodsky:1994kg}%
  \BibitemOpen
  \bibfield  {author} {\bibinfo {author} {\bibfnamefont {Stanley~J.}\
  \bibnamefont {Brodsky}}, \bibinfo {author} {\bibfnamefont {Matthias}\
  \bibnamefont {Burkardt}}, \ and\ \bibinfo {author} {\bibfnamefont {Ivan}\
  \bibnamefont {Schmidt}},\ }\bibfield  {title} {\enquote {\bibinfo {title}
  {{Perturbative QCD constraints on the shape of polarized quark and gluon
  distributions}},}\ }\href {\doibase 10.1016/0550-3213(95)00009-H} {\bibfield
  {journal} {\bibinfo  {journal} {Nucl. Phys. B}\ }\textbf {\bibinfo {volume}
  {441}},\ \bibinfo {pages} {197--214} (\bibinfo {year} {1995})},\ \Eprint
  {http://arxiv.org/abs/hep-ph/9401328} {arXiv:hep-ph/9401328} \BibitemShut
  {NoStop}%
\bibitem [{\citenamefont {Frankfurt}\ \emph {et~al.}(1994)\citenamefont
  {Frankfurt}, \citenamefont {Miller},\ and\ \citenamefont
  {Strikman}}]{Frankfurt:1994hf}%
  \BibitemOpen
  \bibfield  {author} {\bibinfo {author} {\bibfnamefont {L.L.}\ \bibnamefont
  {Frankfurt}}, \bibinfo {author} {\bibfnamefont {G.A.}\ \bibnamefont
  {Miller}}, \ and\ \bibinfo {author} {\bibfnamefont {M.}~\bibnamefont
  {Strikman}},\ }\bibfield  {title} {\enquote {\bibinfo {title} {{The
  Geometrical color optics of coherent high-energy processes}},}\ }\href
  {\doibase 10.1146/annurev.ns.44.120194.002441} {\bibfield  {journal}
  {\bibinfo  {journal} {Ann. Rev. Nucl. Part. Sci.}\ }\textbf {\bibinfo
  {volume} {44}},\ \bibinfo {pages} {501--560} (\bibinfo {year} {1994})},\
  \Eprint {http://arxiv.org/abs/hep-ph/9407274} {arXiv:hep-ph/9407274}
  \BibitemShut {NoStop}%
\bibitem [{\citenamefont {Frankfurt}\ \emph {et~al.}(1993)\citenamefont
  {Frankfurt}, \citenamefont {Miller},\ and\ \citenamefont
  {Strikman}}]{Frankfurt:1993es}%
  \BibitemOpen
  \bibfield  {author} {\bibinfo {author} {\bibfnamefont {L.}~\bibnamefont
  {Frankfurt}}, \bibinfo {author} {\bibfnamefont {G.A.}\ \bibnamefont
  {Miller}}, \ and\ \bibinfo {author} {\bibfnamefont {M.}~\bibnamefont
  {Strikman}},\ }\bibfield  {title} {\enquote {\bibinfo {title} {{Precocious
  dominance of point - like configurations in hadronic form-factors}},}\ }\href
  {\doibase 10.1016/0375-9474(93)90504-Q} {\bibfield  {journal} {\bibinfo
  {journal} {Nucl. Phys. A}\ }\textbf {\bibinfo {volume} {555}},\ \bibinfo
  {pages} {752--764} (\bibinfo {year} {1993})}\BibitemShut {NoStop}%
\end{thebibliography}%


%
\end{document}